\documentclass[useAMS,usenatbib]{mn2e}
\usepackage{amsmath,amssymb,mathrsfs,graphicx,float,latexsym,caption,subcaption}
\usepackage{epstopdf,ragged2e}
\usepackage{longtable}
\usepackage{fixltx2e}
\usepackage{natbib}
\usepackage{hyperref}

\newcommand{\bs}{\boldsymbol{v}}
\newcommand{\bb}{\boldsymbol{B}}
\newcommand{\bsn}{\boldsymbol{\nabla}}
\newcommand{\PLUTO}{{\textsc{\footnotesize{PLUTO}}}}
\newcommand{\ZEUS}{\textsc{\footnotesize{ZEUS}}}
\newcommand{\tA}{\tau_{\textrm{A}}}
\newcommand{\Vm}{V_{\textrm{m}}}
\newcommand{\tC}{\tau_{\textrm{c}}}
\newcommand{\Hm}{H_{\text{max}}}

\def\apj{{ApJ}}

\def\apjs{{The Astrophysical Journal Supplement}}
\def\apjl{{ApJL}}

\def\aap{{A\&A}}

\def\mnras{{MNRAS}}

\def\pasj{{Publications of the Astronomical Society of Japan}}

\def\nat{{Nature}}

\def\prd{{Physical Review D}}

\def\prc{{Physical Review C}}

\def\04a{{2004 a}}
\def\04b{{2004 b}}
\title[Thermal conduction of magnetic mountains]{Relaxation by thermal conduction of a magnetically confined mountain on an accreting neutron star}

\author[A.G.~Suvorov, A.~Melatos] {A.G.~Suvorov$^{1}$\thanks{E-mail:suvorovarthur@gmail.com}, A.~Melatos$^{1,2}$ \\
$^1$ School of Physics, University of Melbourne, Parkville VIC 3010, Australia\\
$^2$ Australian Research Council Centre of Excellence for Gravitational Wave Discovery (OzGrav) \\}

\begin{document}

\date{Accepted ?. Received ?; in original form ?}

\pagerange{\pageref{firstpage}--\pageref{lastpage}} \pubyear{?}

\maketitle

\begin{abstract}

\noindent{A magnetically confined mountain on the surface of an accreting neutron star simultaneously reduces the global magnetic dipole moment through magnetic burial and generates a mass quadrupole moment, which emits gravitational radiation. Previous mountain models have been calculated for idealized isothermal and adiabatic equations of state. Here these models are generalised to include non-zero, finite thermal conduction. Grad-Shafranov equilibria for three representative, polytropic equations of state are evolved over many conduction time-scales with the magnetohydrodynamic solver \textsc{\footnotesize{PLUTO}}. It is shown that conduction facilitates the flow of matter towards the pole. Consequently the buried magnetic field is partially resurrected starting from an initially polytropic Grad-Shafranov equilibrium. The poleward mass current makes the star more prolate, {marginally} increasing its detectability as a gravitational wave source, {though to an extent which is likely to be subordinate to other mountain physics}. Thermal currents also generate filamentary hot spots $(\gtrsim 10^{8} \text{ K})$ in the mountain, especially near the pole where the heat flux is largest, with implications for type I X-ray bursts.} 

\end{abstract}


\section{Introduction} %\label{sec:intro}

Observations of binary neutron stars with white dwarf or supergiant companions and a history of accretion suggest that the neutron star magnetic dipole moment $\mu$ decreases over time, as the accreted mass $M_{a}$ increases \citep{taam86,hue95,zhang06,pat12}. Several theoretical mechanisms exist to explain the trend, such as accelerated Ohmic decay \citep{urp95}, interactions between superfluid vortices and superconductor flux tubes within the stellar interior \citep{srini90}, or the process of magnetic burial \citep{blon86,shibaz}. In magnetic burial, the focus of this paper, matter is guided onto the polar cap by the magnetic field to form a mountain-like density profile supported by the compressed equatorial magnetic field (`magnetic mountain') \citep{bil98,melphi01,pm04,dip12,wang12}. The resulting mass quadrupole moment emits gravitational radiation \citep{usho00,mp05,vig09c,pri11,lasky15}.

% If the mountain induces a filamentary structure in the magnetic field this might allow for fenced-off regions to each provide fuel for thermonuclear runaway events \citep{gal10}. 

%Observation of type I X-ray bursts in advance from low-mass X-ray binaries, which typically have recurrence times of a few hours, allows for additional sensitivity in continuous gravitational wave searches through coherent integration \citep{mp05,gal10,gal14}.

%BURST RECURRENCE TIMES

%While each of these mechanisms can potentially reduce the dipole moment, determining which is most likely to be applicable to generic neutron stars is a largely unsolved problem \citep{melphi01}. We will focus on the magnetic burial scenario within this work, a mechanism which is expected to, in addition to reducing the global dipole moment, excite non-trivial amounts of gravitational radiation \citep{mp05,pri11}. 

The short-term stability and long-term relaxation of a magnetic mountain have been studied by several authors. In the short term, on the Alfv{\'e}n and tearing-mode time-scales, axisymmetric mountain equilibria are susceptible to the undular submode of the Parker instability \citep{pm06,vigstab,vig09} and to pressure-driven toroidal-mode instabilities \citep{cumm01,litwin01,dip13a,dip13b}, once $M_{a}$ exceeds a critical threshold. The system is not necessarily disrupted; the instability saturates, and the mountain adjusts to a new equilibrium, stabilized by magnetic line-tying at the stellar surface and the compressed magnetic `wall' at the equator \citep{vigstab}. In the long term, the mountain relaxes due to Ohmic dissipation \citep{vigohm09}, soft-crust sinking \citep{wette10}, or a combination of the latter two processes \citep{kon02,kon04,kon11}. Its structure is modified also by factors like the Hall effect \citep{cumm04,gv14} and the equation of state (EOS) \citep{pri11}. 

Mountains on recycled pulsars may be responsible for the discrepancy between magnetic field strengths inferred from spin-down and cyclotron line measurements \citep{arons93,nish05}. The local magnetic field can be $\sim 10^{4}$ times stronger than the global value inferred from $\mu$ \citep{mm12,dip12}. Once $M_{a}$ increases beyond a certain level, phase-dependent cyclotron resonance scattering features are predicted to emerge in the X-ray spectrum \citep{pri14}. Additionally, X-ray observations of neutron star binaries reveal type I X-ray bursts with recurrence times ranging between a few minutes and $\sim 10^{3}$ hours \citep{gal08}. Recurrence times $\lesssim 10$ min (e.g. in $4$U $1608$--$522$) are too short for many theoretical ignition models and may indicate the existence of multiple, isolated patches of fuel on the stellar surface \citep{bhat06}, which are fenced-off magnetically if the polar magnetic field geometry is complicated \citep{pmtherm,gal10,misanovic10}. A simultaneous detection of gravitational waves, X-ray bursts with short recurrence times, and cyclotron features in an X-ray binary some time in the future would strongly indicate the presence of a magnetic mountain \citep{hask15}.

In this paper we include thermal conduction in magnetic mountain models self-consistently for the first time. Thermal conduction is potentially important, because the mountain forms at an elevated temperature, caused by accretion-driven heating, and cools through its sides (if accretion is confined to a narrow column) or throughout its volume (once accretion switches off). Thermal fluxes directed out of localized polar hot spots control the instantaneous hydromagnetic structure of the mountain by regulating the EOS \citep{pri11} and the long-term, quasistatic relaxation of the mountain by regulating temperature-sensitive dissipative mechanisms like Ohmic decay \citep{vigohm09}. Modeling thermal conduction in magnetic mountains self-consistently is therefore important for understanding the relationship between hot spots, magnetic fields, X-ray burst activity, and gravitational radiation, providing the basis for multi-messenger tests of the polar magnetic burial scenario.

The purpose of this paper is to elucidate, with the aid of numerical simulations, the dominant thermal processes that modify the short-term structure and long-term evolution of a magnetic mountain, when thermal conduction is ``switched on'' in the model. Predictions are made, in broad qualitative terms, regarding how potentially observable properties (e.g. $\mu$) are affected by thermal conduction. We emphasize, however, that the simulations are not yet at the point where they yield highly realistic mountain models, which are ready to be compared in detail with observational data. Such comparisons would require a more sophisticated description of the stratified structure, composition, and EOS of the crust, better observational knowledge of the high-order magnetic multipoles near the surface, and expanded computational resources to handle the disparate thermal and hydromagnetic time-scales in the problem. Our investigation proceeds in two stages. In Section 2, we use the Grad-Shafranov solver developed by \cite{pm04} and extended by \cite{pri11} to calculate the steady-state structure given an EOS and an initially dipolar magnetic field. We then numerically evolve the Grad-Shafranov equilibrium using the magnetohydrodynamics (MHD) code \PLUTO \,\citep{pluto} with and without thermal conduction in Sections 3 (set-up details and local mountain properties) and 4 (global observables) and compare the effects on potentially observable properties such as $\mu$. Long-term thermal relaxation is explored in Section 5. Finally, the astrophysical implications of the results, including for gravitational wave emission, are discussed briefly in Section 6.

%a series of hydrodynamical simulations building on the work of \cite{pm04} (henceforth PM04) and \cite{pri11} (henceforth P11). Using the previously developed quasi-static Grad-Shafranov code developed in PM04 and P11 we solve for the static mountain structure given an EOS for the infalling matter and a natal dipolar magnetic field for the host neutron star (described in Sec. 2). We then numerically evolve the mountains in time using the unimodular code \PLUTO \,\citep{pluto} with and without thermal conduction to explore the resulting time-dependent equilibria (Sec. 2 \& 3). 

\section{Polar magnetic burial}

\subsection{Qualitative behaviour}

\begin{figure}
%\centering
\includegraphics[width=0.473\textwidth]{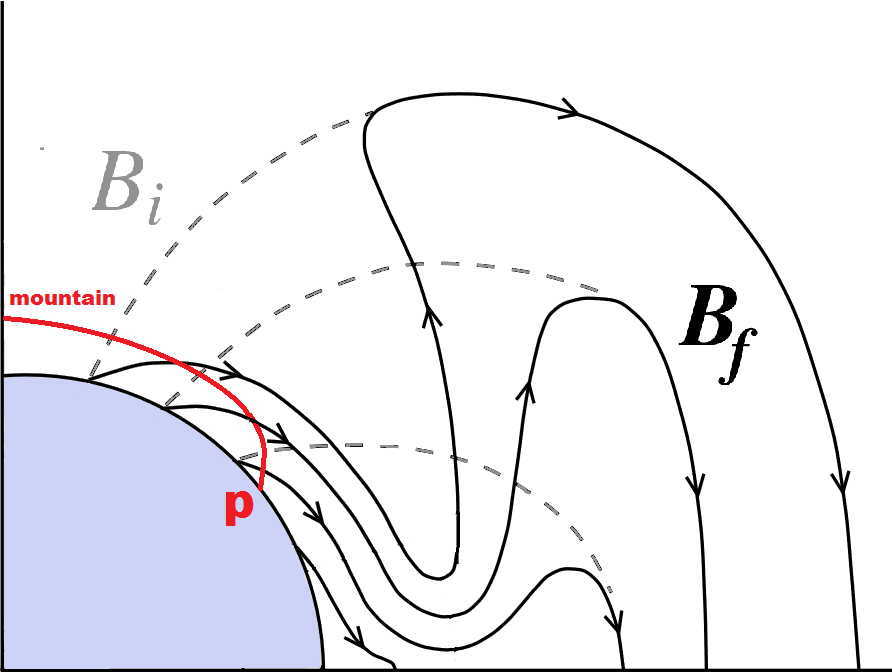}
\caption{Schematic diagram (not to scale) of the magnetic field lines prior to (dashed curves) and after (solid curves) accretion. The shaded region represents the neutron star. The mountain fills the region bounded between the red curve and the stellar surface. The pressure gradient at the base of the mountain is balanced by the Lorentz force in the compressed, equatorial magnetic belt. The edge of the mountain (point $p$) moves towards the equator, as $M_{a}$ increases. [Adapted from Melatos \& Phinney (2001).] \label{mountdiag}
}
\end{figure}

During accretion, the neutron star's polar magnetic field buckles underneath the infalling matter, and the field lines spread equatorially due to flux freezing. The lateral pressure gradient at the base of the accreted mountain is balanced by the Lorentz force in the compressed, equatorial magnetic belt. This process is illustrated schematically in Figure \ref{mountdiag} [see also Figure 6 of \cite{pri11}]. The compressed magnetic field is more intense than the pre-accretion field locally, due to magnetic flux conservation, but the global moment $\mu$ reduces, because the magnetic distortion induces screening currents, which reduce the radial magnetic field near the pole \citep{vigstab,mm12}. %Gravitational radiation is excited during and after burial, because the mountain is inherently non-spherical and therefore induces a density anisotropy into the mountain plus neutron-star system.

It is observed that $\mu$ decreases with $M_{a}$ in binary systems \citep{taam86,hue95,zhang06}. \cite{shibaz} proposed the widely used, empirical law 
\begin{equation} \label{eq:empirical}
\mu = \mu_{i} \left( 1 + M_{a} / M_{c} \right)^{-1} .
\end{equation}
In \eqref{eq:empirical}, we define $M_{c}$ to be the critical accreted mass, for which the global dipole moment is halved. The dipole moment before accretion begins is given by $\mu_{i} = B_{\star} R_{\star}^{3}$. We take $R_{\star} = 10^{6} \text{ cm}$ for the stellar radius and $B_{\star} = 10^{12.5} \textrm{ G}$ for the natal magnetic field strength {at the polar surface}, in line with population synthesis models \citep{arzo,kaspi}. Self-consistent, MHD simulations reproduce the empirical scaling \eqref{eq:empirical} for small accreted masses $M_{a} \ll M_{c}$ in isothermal and adiabatic mountains with $M_{c} \sim 10^{-6} M_{\odot}$ and $M_{c} \sim 10^{-8} M_{\odot}$ respectively \citep{pm04,vig09,pri11}. For $10^{-1} \lesssim M_{a}/ M_{c} \lesssim 10$, the simple estimate in \eqref{eq:empirical} breaks down and the burial effect is better represented by a power-law $\mu / \mu_{i} = (M_{a} / M_{c})^{-a}$, where $1 \leq a \leq 2.47$  depends on the EOS [see section 4.1 of \cite{pri11} and Fig. 8(c) of \cite{pm04}]. Numerical difficulties prevent simulations from probing the regime $M_{a} / M_{c} \gtrsim 10$, where a significant deviation from \eqref{eq:empirical} is expected \citep{hask15}, though Ohmic diffusion sets a burial limit of $\mu / \mu_{i} \gtrsim 10^{-8}$ \citep{vigohm09}.

%is corrected in the form $\mu = \mu_{i} \left( 1 + M_{a} / M_{c} \right)^{-2.25 \pm 0.22}$.
%[2-3 sentences about how GS papers relate to estimate (1); vigelius, payne, etc. Numbers, refs.]

The critical mass $M_{c}$ depends strongly on the EOS \citep{pri11}. For a softer EOS, the mountain has a relatively small thickness $(\Hm \sim 10^{3} \text{ cm})$, because the material is easier to compress. Strong local magnetic fields $(\lesssim 10^{15} \text{ G})$ exist near the stellar surface, as polar field lines buckle, and the polar magnetic flux is squeezed into a small volume. Consequently, the screening currents flow closer to the stellar surface for a softer EOS than for a harder EOS, and $\mu$ reduces less for a given $M_{a}$. For a softer EOS, the critical mass is found to lie in the range $10^{-5} \lesssim M_{c} / M_{\star} \lesssim 10^{-2}$ \citep{pm04}. For a harder EOS, the mountain is thicker $(\Hm \sim 10^{4} \text{ cm})$, and $\mu$ reduces further. For a polytropic EOS with index $\Gamma \gtrsim 4/3$, the critical mass is found to lie in the range $10^{-9} \lesssim M_{c} / M_{\star} \lesssim 10^{-6}$ \citep{pri11}. \cite{pri11} found $M_{c} \propto B_{\star}^2$. %The proportionality constant $k_{\Gamma}$, relating the pressure $p$ and density $\rho$ in $p = k_{\Gamma} \rho^{\Gamma}$, typically decreases as $\Gamma$ increases. %in the range $1 \leq \Gamma \leq 5/3$.}

%[Add 2-3 sentences on Priymak results for mu vs Ma and Mc vs B.]

%Isothermal (soft) and polytropic} (hard) EOS represent two extremes.

%For simplicity we consider two models of accreted plasma; isothermal (soft, model A) and isentropic, classical, ideal gasses (hard, models B and C) with index $\gamma = 5/3$, for two different proportionality constants (see Table \ref{tab:table1eos}).

%The self-consistent equilibrium determined by the EOS is found through a sequence of quasi-static Grad-Shafranov evolutions, and is thus insensitive to time-dependent processes such as thermal conduction.
 
%By including thermal conduction in the mountain model --- the aim of this paper --- we can evolve the mountain dynamically between the two extremes and quantify the results in terms of the quasi-static change in $M_{c}$. 

%}

%[RE THIS -- explain that we use these as an approximation and that TC should matter]:
%They also correspond to the limits of infinite and zero thermal conduction, respectively. 

The mountain mass quadrupole moment can be expressed in terms of the mass ellipticity  $\epsilon$, which is given approximately by \citep{mp05}
\begin{equation} \label{eq:empepsilon}
\epsilon \approx \left( 5 M_{a} / 4 M_{\star} \right) \left( 1 + 9 M_{a} / 8 M_{c} \right)^{-1} ,
\end{equation}
%and
%\begin{equation}
%\mu \approx B_{\star} R_{\star}^{3} \left( 1 + 3 M_{a} / 4 M_{c} \right)^{-1},
%\end{equation}
where $M_{\star} = 1.4 M_{\odot}$ is the stellar mass. Therefore, for a given accreted mass, there is a one-to-one relationship between $\mu$ and $\epsilon$ through \eqref{eq:empirical} and \eqref{eq:empepsilon} which depends on the value of $M_{c}$. The mass quadrupole moment emits gravitational radiation, as the star spins \citep{mp05}. The implications are discussed in Section 4.2.

%\begin{multicols}{2}
%\end{multicols}

\subsection{Hydromagnetic equilibrium}

The hydromagnetic structure of the mountain has been calculated for various EOS previously \citep{pm04,pm06,vigstab,pri11,dip13a}. In this paper, we start by solving for the structure of a steady-state $(\partial / \partial t = 0)$ and immobile ($\bs = 0$, where $\boldsymbol{v}$ is the fluid velocity) mountain. We then input the result into \PLUTO \,as the starting point for time-dependent simulations which include thermal conduction. In the first stage we do not model growth of the mountain but rather solve for a self-consistent equilibrium (i.e. hydromagnetic force balance) given a certain amount of accreted mass and an EOS. Time-dependent simulations in the literature confirm that the equilibrium agrees closely with mountains built from scratch by injecting mass from below \citep{vig09c,wette10}.

We assume that the magnetic field $\bb$ may be described through an axisymmetric Chandrasekhar decomposition without a toroidal component\footnote{ Three-dimensional mountain equilibria for $M_{a} \gtrsim M_{c}$ are susceptible to Parker-like \citep{vigstab} and ballooning \citep{dip13b} instabilities with EOS-dependent growth rates \citep{kos06}. However, the instabilities do not disrupt the mountain; they reduce the ellipticity by $\lesssim 30$ percent when they saturate \citep{vigstab}. Furthermore, previous three-dimensional, time-dependent simulations reveal that the magnetic field relaxes to an almost axisymmetric configuration within a few Alfv{\'e}n times \citep{pstab07,vigstab}.} for simplicity, i.e. $\boldsymbol{B}$ takes the form
\begin{equation} \label{eq:chandra}
\bb = \frac {\bsn \psi} {r \sin \theta} \times \hat{\boldsymbol{e}}_{\varphi}
\end{equation}
in spherical coordinates $\left(r,\theta,\varphi\right)$, where $\psi(r,\theta)$ is a scalar flux function \citep{C56}. The mountain equilibrium is determined by the force balance (Grad-Shafranov) equation \citep{pm04,dip12}
\begin{equation} \label{eq:gradshaf}
\bsn p + \rho \bsn \phi + \Delta^2 \psi \bsn \psi = 0,
\end{equation}
where $\Delta^2$ denotes the Grad-Shafranov operator,
\begin{equation} \label{eq:gsoperator}
\Delta^2 = \frac {1} {4 \pi r^2 \sin^2\theta} \left[ \frac {\partial^2} {\partial r^2} + \frac {\sin \theta} {r^2} \frac {\partial} {\partial \theta} \left( \frac {1} {\sin \theta} \frac {\partial} {\partial \theta} \right) \right],
\end{equation} 
$p$ is the fluid pressure, $\rho$ is the mass density, and $\phi$ is the gravitational potential. For now, we assume that the accreted matter obeys a barotropic EOS, $p = p(\rho)$, once equilibrium is reached. Although neutron stars are expected to be non-barotropic \citep{reissen1}, the barotropic and non-barotropic solutions to the Grad-Shafranov problem are broadly similar, even in magnetars \citep{mmra11,MSM15}.

%previous calculations have suggested that there is little quantitative difference in global observables between barotropic and non-barotropic solutions to the Grad-Shafranov problem, even for magnetars \citep{mmra11,MSM15}.

For $p = p(\rho)$, equation \eqref{eq:gradshaf} implies that $\rho$ and $\psi$ satisfy
\begin{equation} \label{eq:somereln}
0 = \frac {d p} {d \rho} \left( \frac {\partial \rho} {\partial r} \frac {\partial \psi} {\partial \theta} - \frac {\partial \rho} {\partial \theta} \frac {\partial \psi} {\partial r} \right) + \rho \left( \frac {\partial \phi} {\partial r} \frac {\partial \psi} {\partial \theta} - \frac {\partial \phi} {\partial \theta} \frac {\partial \psi} {\partial r} \right).
\end{equation}
Equation \eqref{eq:somereln} can be solved exactly using the Lagrange-Charpit method to yield \citep{lagchap}
\begin{equation} \label{eq:lagcha}
\int \frac {d p} {d \rho} \frac {d \rho} {\rho} = F(\psi) - \left( \phi - \phi_{0} \right),
\end{equation}
where $\phi_{0}$ denotes a reference gravitational potential at the neutron star surface, and $F$ is an arbitrary function of the scalar flux. Given $p = p(\rho)$, the integral in \eqref{eq:lagcha} can be evaluated, in principle, to express $\rho$ in terms of $\psi$ or vice-versa. 

In order to obtain a one-to-one correspondence between the pre- and post-accretion states that respects flux-freezing, we demand that the steady-state, mass-flux ratio $d M / d \psi$, defined as the mass enclosed between the infinitesimally separated flux surfaces $\psi$ and $\psi + d \psi$, equals that of the initial state plus any accreted matter \citep{alf43,mous74,melphi01}.  This restriction on $M(\psi)$ leads to the constraint \citep{pm04}
\begin{equation} \label{eq:fluxfreezing}
\frac {d M} {d \psi} = 2 \pi \int_{C} ds \rho \left[ r(s), \theta(s) \right] r \sin \theta | \bsn \psi |^{-1} ,
\end{equation}
where $C$ is the curve $\psi[r(s),\theta(s)] = \psi$ parametrized by the arc length $s$. Equation \eqref{eq:fluxfreezing} can be solved by inverting \eqref{eq:lagcha} for $\rho$ in terms of $\psi$ given $p=p(\rho)$, leading to a unique expression for the function $F(\psi)$ given $M(\psi)$. The explicit forms of $F$ for adiabatic and isothermal EOS can be found in \cite{pri11} and \cite{pm04} respectively.

We solve \eqref{eq:gradshaf} simultaneously with \eqref{eq:fluxfreezing} numerically using the relaxation algorithm described in \cite{pm04} and later extended by \cite{pri11}. Specifically, the solver employs iterative under-relaxation combined with a finite-difference Poisson solver to solve \eqref{eq:gradshaf} for $\psi$, obtain $\rho$ from \eqref{eq:fluxfreezing}, and feed the result back into \eqref{eq:gradshaf} iteratively, until convergence is achieved. Additional information regarding units, convergence, and stability can be found in the aforementioned papers and is not repeated here; see also \cite{pstab07} and \cite{vigstab}. In accord with previous work, we prescribe the mass-flux distribution in one hemisphere to be \citep{mp05}
\begin{equation} \label{eq:massfluxexp}
M(\psi) = \frac {M_{a} \left[ 1 - \exp\left(-\psi/\psi_{a}\right)\right]} {2 \left[ 1 - \exp\left(-b\right)\right]},
\end{equation}
where $M_{a}$ is the accreted mass, $\psi_{a}$ labels the field line at the polar-cap boundary (that closes just inside the inner edge of the accretion disc), and we define $b = \psi_{\star} / \psi_{a}$, where $\psi_{\star}$ labels the total hemispheric flux. Throughout this paper we set $b=3$ to ensure numerical stability.

For simplicity we assume a constant gravitational acceleration, with $\phi(r) = G M_{\star} r / R_{\star}^2$, where $R_{\star}$ is the stellar radius, and make the Cowling approximation (i.e. we ignore self-gravity). These assumptions are justified, because the mountain never rises more than $\sim 10^{4}$ cm above the surface at $r = R_{\star}$, and we consider systems with $M_{a} / M_{\odot} \lesssim 10^{-1}$ [see section 2.1 of \cite{pri11}]. Additionally, \cite{haskell06} and \cite{yoshida13} found that the Cowling approximation alters the mass ellipticity by at most a factor of $\sim 3$ even for the strongest magnetar fields $(\lesssim 10^{16} \text{ G})$.

Equation \eqref{eq:gradshaf} is solved subject to physically motivated boundary conditions, which carry through to the evolution experiments in \PLUTO \,in Sec. 2.4 (see also Appendix A). Following previous work, we set $\psi(R_{\textrm{in}},\theta) = \psi_{\star} \sin^2\theta$ (surface dipole), $\partial \psi / \partial r (R_{\textrm{m}},\theta) = 0$ (Neumann outflow\footnote{{Ideally, this condition would be replaced by a dipolar field condition at the outer edge, i.e. $\psi(R_{\text{m}},\theta) = \psi_{\text{m}} \sin^2\theta$ for some value of $\psi_{\text{m}}$, so as not to introduce artificial magnetic multipoles (including a monopole) far from the stellar surface. However, in order to assign a value to $\psi_{\text{m}}$, as necessary for numerical computation, we need to know by how much magnetic burial reduces the dipole moment, with $\mu / \mu_{i} \propto \psi_{\text{m}} / \psi_{i}$. In principle, it is possible to adjust $\psi_{\text{m}}$ iteratively in order to obtain a self-consistent simulation, but this is technically challenging (see also Footnote 4). A thorough discussion of the issue can be found in Sec. 4.3 of \cite{vigstab} and in Sec. 4.1 of \cite{pstab07}, as well as references therein. In particular, \cite{vigstab} showed that the density distribution (Fig. 14 of the latter reference, left panel) is virtually indistinguishable between Neumann and dipole condition equilibria, while the magnetic field lines (right panel) tend to agree except in the outermost regions, where the plasma density is low. The Neumann outflow condition artificially increases $\mu$, relative to corresponding \ZEUS \,equilibria, by $\lesssim 10 \%$ [see Fig. 3(f) of \cite{pstab07}].}}), $\psi(r,0) = 0$ (straight polar field line), and $\partial \psi / \partial \theta (r , \pi/2 ) = 0$ (equatorial symmetry), where $R_{\textrm{in}} \leq r \leq R_{\textrm{m}}$ and $0 \leq \theta \leq \pi/2$ demarcate the computational volume \citep{pm04,mp05,vigstab,pri11}. The outer radius $R_{\textrm{m}}$ is chosen large enough to encompass all the screening currents and the outer edge of the accreted matter; we set $R_{\textrm{m}} = 1.4 R_{\star}$ throughout this paper. The boundary conditions on $\psi$ are reformulated as conditions on $\bb$ through \eqref{eq:chandra} and conditions on $\rho$ (and hence $p$) through \eqref{eq:fluxfreezing}. In the time-dependent \PLUTO \,simulations (see Sec. 2.4) we also stipulate no slip at $R_{\text{in}}$, outflow at $R_{\text{m}}$, and reflecting boundary conditions on $\boldsymbol{v}$ at the equator.

\begin{figure}
\includegraphics[width=0.473\textwidth]{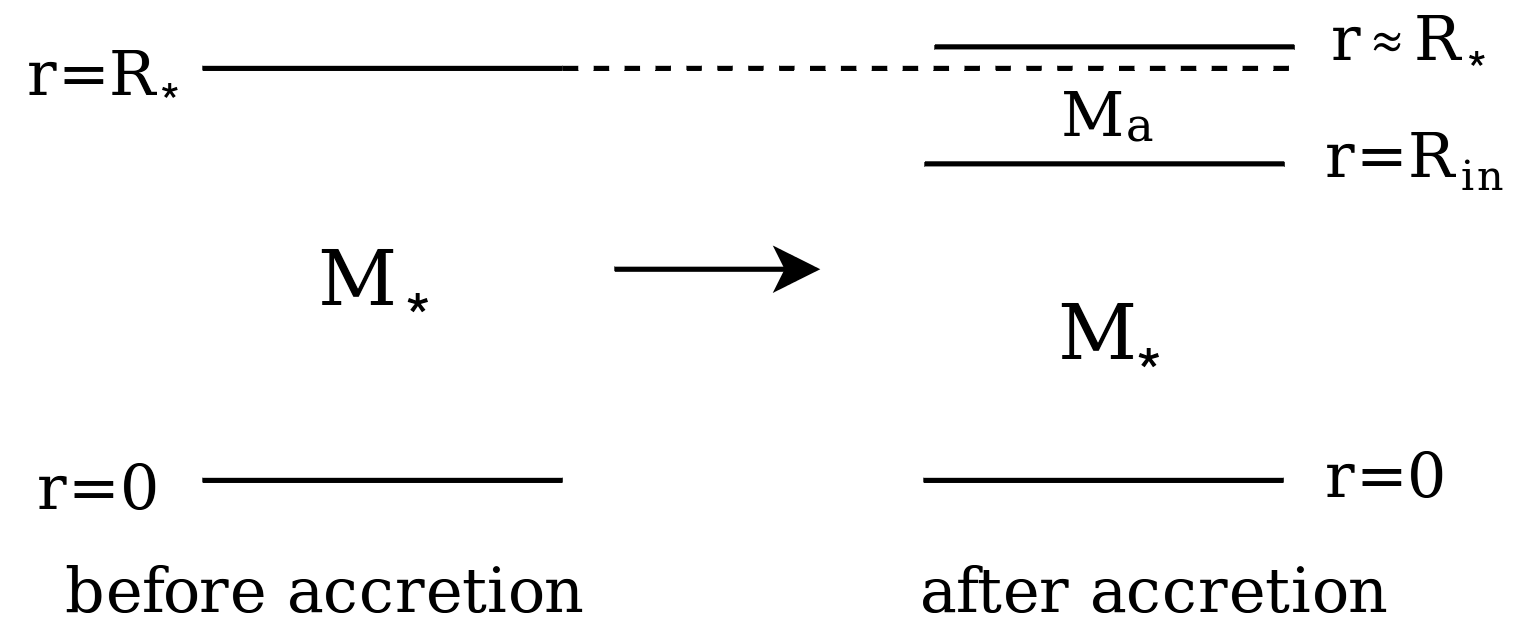}
\caption{Schematic diagram (not to scale) of the radial computational setup, described in Sec. 2.2, of an accreted crust on a neutron star. The inner computational boundary $R_{\text{in}}$ is determined by $M_{a} = \int^{r=R_{\star}}_{r=R_{\text{in}}} d^{3} \boldsymbol{x} \, \rho$. If sinking is included in the model (see Sec. 2.3), the mountain height reduces on the hydrostatic settling time-scale. \label{compsetup}
}
\end{figure}

Although $R_{\textrm{in}}$ is treated as a hard surface for simplicity, it is not so in reality; a mountain several tens of meters high, whose base reaches neutron drip densities, sinks into the lower-density substrate \citep{wette10,pri11}. A full treatment of sinking requires time-dependent simulations. \cite{wette10} showed that the results are approximated reasonably by hard-surface solutions {(the mountain ellipticity decreases by a factor $\lesssim 2$ for soft crust solutions relative to hard-surface solutions)}, if $R_{\textrm{in}}$ corresponds to the layer above which the mass equals $M_{a}$; i.e. one has $R_{\textrm{in}} < R_{\star}$ and $M_{a} = \int^{r=R_{\star}}_{r=R_{\textrm{in}}} d^{3} \boldsymbol{x} \, \rho$. As $R_{\textrm{in}}$ is fixed, the stellar radius $R_{\star}$ varies slightly ($\leq 1$ per cent for $M_{a} \leq 10^{-4} M_{\odot}$) between models with different $M_{a}$ but the same EOS. Figure \ref{compsetup} visually demonstrates the relationship between $R_{\text{in}}, R_{\star}$, and $M_{a}$. 

\subsection{Equation of state}

Crustal matter experiencing compression due to accretion undergoes a variety of non-equilibrium nuclear processes, such as electron captures and beta decay, neutron emission and absorption, and pyconuclear fusion, all of which play a role in determining the EOS of the accreted crust \citep{sato79,miralda90,cham08}. The original outer crust, consisting of cold, catalysed matter, is replaced by a new, non-catalysed crust after $\lesssim 10^{5} \text{ yr}$ \citep{haen89}. The EOS of an accreted, non-catalysed crust, relevant for our calculations, has been calculated numerically by \cite{haen90} by modeling the non-equilibrium processes listed above, using the compressible liquid drop model of \cite{mackie77} to estimate the various thermodynamic rates which feed into the Gibbs equation [see also Sec. 2.4 of this paper and \cite{kogan79}].

%{ Thermal fluxes also modify the polytropic index \citep{LL59}; see Sec. 2.4. }
% which influences the EOS as heat is transported over time; cf \cite{wette10}.

In this paper we consider three idealized yet physically motivated polytropic EOS: a single-index model which best approximates (in a least-squares sense; see below) a realistic accreted crust (model A), and two of the classical ideal gas models considered by \cite{pri11}, corresponding to a gas of non-relativistic degenerate electrons (model B), and a gas of non-relativistic degenerate neutrons (model C). Their parameters are summarised in Table \ref{tab:table1eos}; see also \cite{pri11}. Models A, B, and C apply only at $t=0$; they are used to construct initial conditions for \PLUTO. The Grad-Shafranov solver developed by \cite{pri11} quasi-statically determines the `end-state' of an adiabatic accretion process in the absence of thermal conduction. As noted above, we initialise \PLUTO \,with an $M_{a}$-dependent Grad-Shafranov equilibrium to avoid numerical difficulties, cf. \cite{wette10}. In reality, however, the true end-state of accretion depends on $\dot{M}_{a}$ as well as $M_{a}$. Accreted plasma on the stellar surface is expected to be approximately isothermal ($\Gamma \approx 1$) at all depths for low-accretion rates $\dot{M}_{a} \lesssim 10^{-10} M_{\odot} \text{ yr}^{-1}$ \citep{fuji84,zdun92}. In contrast, a crust formed on a star accreting near the Eddington limit $\dot{M}_{a} \gtrsim 10^{-8} M_{\odot} \text{ yr}^{-1}$ has a more complicated polytropic EOS with a depth-dependent adiabatic index ($1 \lesssim \Gamma \lesssim 5/3$) \citep{bil98,brown00}. Hence the time-dependent accretion process, which is not modeled here except implicitly through \eqref{eq:massfluxexp}, directly affects the softness or hardness of the EOS. In particular, a self-consistent model of accretion along the lines of the sinking problem treated by \cite{wette10} would lead to end-state values of $k_{\Gamma}$ and $\Gamma$ which depend on both $M_{a}$ and $\dot{M}_{a}$ \citep{fuji84,rutledge}. The numerical experiments we conduct in \PLUTO, which evolve the EOS via thermal conduction, partially account for the effects of a near-Eddington accretion rate on the crustal EOS \emph{a posteriori} (see Sec. 2.4). This procedure has been validated in the absence of thermal conduction by \cite{wette10}. %for which a non-catalysed crust with $\Gamma \gtrsim 1$ acts an appropriate initial condition \citep{zdun92,brown00}, through thermal conduction  

\begin{figure*}
\includegraphics[width=\textwidth]{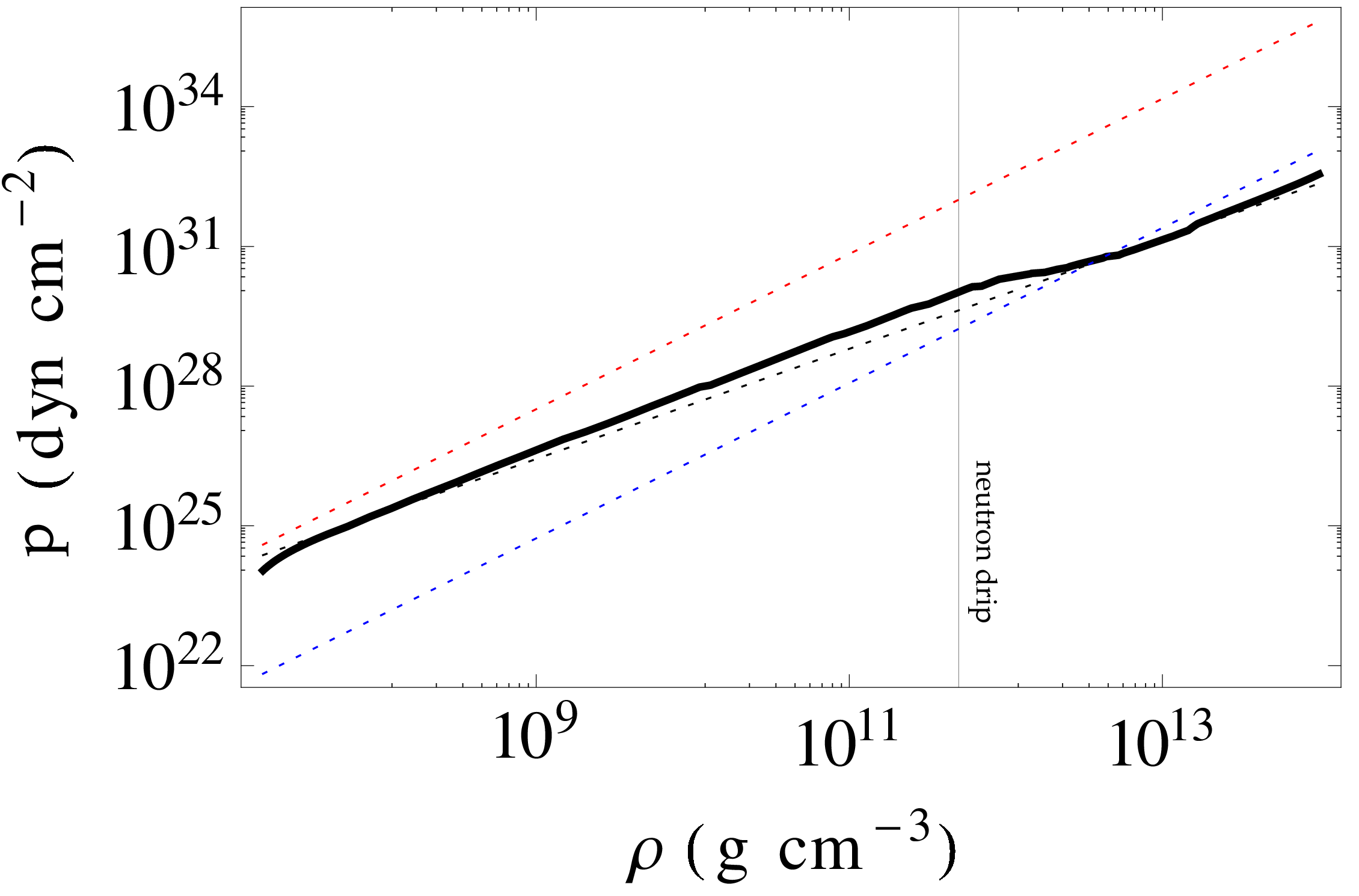}
\caption{Realistic EOS of a non-catalysed, accreted crust (Haensel \& Zdunik 1990b; Douchin \& Haensel 2001) (solid curve) together with the approximate, single-index EOS used in this paper: model A (black dashed curve), model B (red dashed curve), and model C (blue dashed curve). For reference, the neutron drip density [$\sim 5 \times 10^{11} \text{ g cm}^{-3}$ in an accreted crust (Chamel et al. 2015)] is shown by a vertical line. \label{zdunik}
}
\end{figure*}

In Figure \ref{zdunik} we graph pressure-density relationships for models A, B, and C (broken curves) together with the numerical results of \cite{haen90} (solid curve). For $\rho \gtrsim 1.5 \times 10^{13} \text{ g cm}^{-3}$, the maximum density computed by \cite{haen90}, we graph the inner-crust model of \cite{douchin01}, also computed using the compressible liquid drop model \citep{mackie77}. Denoting the neutron drip density by $\rho_{\text{nd}}$ [$\rho_{\text{nd}} \sim 5 \times 10^{11} \text{ g cm}^{-3}$ in an accreted crust \citep{cham15}], we see that the numerical results are approximated adequately by models B and C used in previous work \citep{pri11} in the regimes $\rho \ll \rho_{\text{nd}}$ and $\rho \gtrsim \rho_{\text{nd}}$ respectively. On the other hand, model A is constructed to uniformly approximate the realistic EOS for all $\rho \lesssim 10^{14} \text{ g cm}^{-3}$. The parameters $k_{\Gamma}$ and $\Gamma$ for model A are computed by fitting $p(\rho) = k_{\Gamma} \rho^{\Gamma}$ with the Levenberg-Marquardt (damped least-squares) algorithm \citep{levmac} to the data collated in Table 1 of \cite{haen90}. Denoting the \cite{haen90} numerical pressure by $p^{\text{HZ}}$ and the model A pressure by $p^{\text{A}}$, the fit yields relative errors of $0.94 \leq p^{\text{HZ}}/p^{\text{A}} \leq 2.51$ for $10^{8} \leq \rho/ \text{ g cm}^{-3} \leq 10^{14}$. Throughout most of the mountain volume by mass, i.e. for $\rho \geq 10^{12} \text{ g cm}^{-3}$, the errors drop to $\leq 6$ percent, with $0.95 \leq p^{\text{HZ}}/p^{\text{A}} \leq 1.06$.  For $\rho \leq 10^{9} \text{ g cm}^{-3}$ we have $0.94 \leq p^{\text{HZ}}/p^{\text{A}} \leq 1.07$.

%\begin{multicols}{2}
%\end{multicols}
%as a linear interpolation, actual data shown in triangles

\begin{table}
%\begin{minipage}{\textwidth}
\centering
% \begin{minipage}{160mm}

  \caption{EOS parameters for numerical mountain models. We assume a polytropic, single-index EOS with $p(\rho) = k_{\Gamma} \rho^{\Gamma}$, where $k_{\Gamma}$ is measured in cgs units (dyn $\text{ g}^{-\Gamma}$ $\text{ cm}^{3 \Gamma -2 }$) (Shapiro \& Teukolsky 1983; Priymak et al. 2011). }
  \begin{tabular}{lccc}
  \hline
Model & $k_{\Gamma}$ (cgs)  & $\Gamma$  & EOS  \\
\hline
A &   $6.18 \times 10^{15}$  & $1.18$ & Realistic accreted crust   \\
B   &  $3.16 \times 10^{12}$ & $5/3$ & Isentropic gas; degenerate $e^{-}$   \\
C   & $ 5.38 \times 10^{9}$  & $5/3$ & Isentropic gas; degenerate $n$ \\

\hline
\end{tabular}
%\end{minipage}
\label{tab:table1eos}
%\end{minipage}
\end{table}

A piecewise polytropic fit to the solid curve in Fig. \ref{zdunik} (e.g. a spline fit to $\log p$ vs $\log \rho$) is a better approximation than the uniform, $\Gamma = 1.18$ fit in model A. As a practical matter, however, it is difficult to generalise the Grad-Shafranov calculation in Sec. 2.2, especially the explicit formula for $F(\psi)$ [equation (8) in \cite{pri11}], to apply to multiple layers with proper matching at the inter-layer boundaries. As the Grad-Shafranov calculation is an essential input to the \PLUTO \,simulations, the uniform $\Gamma = 1.18$ approximation is the best we can do for now. For this reason, among others, the final results should be viewed as qualitatively representative models of the thermal conduction physics rather than quantitatively accurate mountain models to be compared in detail to observational data.

Different EOS predict different maximum (base) densities $\rho_{\text{max}}$ and heights $H$ for any given $M_{a}$. In Table \ref{tab:tsinking} we list the characteristic $\rho_{\text{max}}$ and $H$ for runs performed in this paper (see Sec. 3.2) together with a {rough estimate for the} expected depth within a neutron star to which $\rho_{\text{max}}$ corresponds. {Note that the sinking depths listed in Table \ref{tab:tsinking} apply for stellar mass $M_{\star} = 1.4 M_{\odot}$, and a Skyrme-type EOS at zero temperature, used to describe both the crust and the liquid core, based on the effective nuclear interaction SLy \citep{douchin01}. Different EOS and stellar masses lead to different sinking depths.} For models A and C, the theoretical depth exceeds the simulated height of the mountain. Depending on the crustal elasticity \citep{cham08}, this indicates that mountain matter should sink beneath the surface and influence the hydromagnetic structure of the star \citep{kon02}. However, using breaking strain arguments, it has been shown that realistic crustal ellipticites of neutron stars cannot exceed $\sim 10^{-6}$ \citep{haskell06,mcdaniel13}, which is less than those associated with accreted mountains (see Sec. 4.2). As such, any gravitational radiation due to crustal quadrupole moment generation via back-reaction effects from a sinking mountain is likely to be dwarfed by the radiation due to the mountain itself \citep{wette10}, though there may be interesting consequences for other phenomena, e.g. crust-core coupling \citep{glam06}. {The lateral $(\theta)$ structure of the mountain is not affected greatly by sinking, as shown by \cite{wette10}. Hence the main effect of sinking on $\epsilon$ is to reduce it by a factor $\approx \left(R_{\text{pre}} / R_{\text{post}}\right)^{5}$, where $R_{\text{pre}}$ and $R_{\text{post}}$ are the characteristic radii of the base of the mountain before and after sinking, respectively, and ``before sinking" here means ``in the context of a hard-surface Grad-Shafranov calculation''. In any case, because we do not model sinking, the values of the ellipticities (and heights) presented in this paper should be taken as upper limits.} Modeling a realistic neutron star together with a sinking mountain in a way that simultaneously tracks the Alfv{\'e}n and sinking time-scales is a difficult problem that will be considered in future work.

\begin{table}
%\begin{minipage}{\textwidth}
\centering
% \begin{minipage}{160mm}

  \caption{Characteristic maximum density, mountain height, and {approximate} depth beneath the surface where the pre-accretion density is $\rho_{\text{max}}$ (i.e. characteristic sinking depth),  {estimated for a neutron star of mass $M_{\star} = 1.4 M_{\odot}$ with a Skyrme EOS}. }
  \begin{tabular}{lccc}
  \hline
Model & $\rho_{\text{max}} (\text{g cm}^{-3})$ & Height (cm) & Sinking depth (cm) \\
\hline
A &   $2 \times 10^{13} $  & $3 \times 10^{3}$ & $\sim 10^{4} - 10^{5}$\\
B   & $8 \times 10^{8}$ & $4 \times 10^{4}$ & $\sim 10^{3} - 10^{4}$   \\
C   & $3 \times 10^{11} $  & $5 \times 10^{3}$ & $\sim 10^{4} - 10^{5}$ \\

\hline
\end{tabular}
%\end{minipage}
\label{tab:tsinking}
%\end{minipage}
\end{table}

\subsection{MHD evolution}
The steady-state solution to the Grad-Shafranov problem in section 2.2 serves as initial data for evolving the mountain dynamically. In the absence of viscosity and under the assumptions of infinite electric conductivity\footnote{{The Ohmic diffusion and thermal conduction time-scales (see Sec. 2.5) are in the ratio $\tau_{\text{d}} / \tC \approx 2 \times 10^{-14}  \left( \sigma / \text{ s}^{-1} \right) |\bb| / |\bsn^2 \bb | L^{-2} \left( \rho / 10^{10}  \text{ g cm}^{-3} \right)^{-1}$, for characteristic length-scale $L$ and electrical conductivity $\sigma$. In the crust-magnetosphere interface, one has $\sigma \lesssim 10^{16} \text{ s}^{-1}$ \citep{akgun18}. In the inner crust one has $\sigma \gtrsim 10^{24} \text{ s}^{-1}$ \citep{pot99,pot13}. Hence we find $\tau_{\text{d}} / \tC \gg 1$ throughout the computational volume for the range of accreted masses considered in this paper, even in regions with strong magnetic gradients, because the density is low there $\left(\rho \lesssim 10^{10} \text{ g cm}^{-3}\right)$. We can therefore safely ignore the effects of Ohmic diffusion over the time-scales simulated within this paper; see also \cite{vig09c}.}} (ideal MHD) and the Cowling approximation, the evolution is governed by the continuity, Euler, and Faraday equations, which read \citep{LL59}
\begin{equation} \label{eq:cty}
0 = \frac {\partial \rho} {\partial t} + \bsn \cdot \left( \rho \bs \right),
\end{equation}
\begin{align} \label{eq:euler}
%\begin{aligned}
0 =& \rho \frac {\partial \bs} {\partial t} + \rho \left( \bs \cdot \bsn \right) \bs + \bsn p + \rho \bsn \phi \nonumber \\ 
& -   \left( 4 \pi \right)^{-1} \left[ \left(\bsn \times \bb\right) \times \bb \right] ,
%\end{aligned}
\end{align}
and
\begin{equation} \label{eq:faraday}
0 = \frac {\partial \bb} {\partial t} - \bsn \times \left( \bs \times \bb \right),
\end{equation}
%\begin{equation} \label{eq:poisson}
%0 = \bsn^2 \phi - 4 \pi G \rho,
%\end{equation}
respectively, given $\phi$. The MHD equations \eqref{eq:cty}--\eqref{eq:faraday} are closed by the energy equation,
\begin{align} \label{eq:energy}
%\begin{aligned}
\bsn \cdot \boldsymbol{F} =& \bsn \cdot \left[ \left( \varepsilon + \frac {\rho \bs^2} {2} + p + \frac {\bb^2} {4 \pi} + \rho \phi \right) \bs  - \frac {\bb \left( \bs \cdot \bb \right)} {4 \pi} \right] \nonumber \\
&+\frac {\partial} {\partial t} \left( \varepsilon + \frac {\rho \bs^2} {2} + \frac {\bb^2} {8 \pi} + \rho \phi \right),
%\end{aligned}
\end{align}
where $T$ is the temperature, $\boldsymbol{F}$ is the heat flux, and $\varepsilon$ is the internal energy \citep{ST83}. The kinetic properties of the fluid determine the internal energy in terms of the thermodynamic variables $p, \rho$, and temperature $T$, i.e. $\varepsilon = \varepsilon(p,\rho,T)$ \citep{kundu}. The Gibbs fundamental equation,
\begin{align}
0 &= T d S - p d V - d \varepsilon \label{eq:gibbs1} \\
&= T d S - p d V - \left( \frac {\partial \varepsilon} {\partial p} dp + \frac {\partial \varepsilon} {\partial \rho} d\rho + \frac {\partial \varepsilon} {\partial T} dT \right),  \label{eq:gibbs}
\end{align}
where $V$ is the system volume and $S$ is the entropy, provides an additional constraint for the state variables. We then have seven scalar equations, namely \eqref{eq:cty}--\eqref{eq:gibbs1}, for seven variables in the axisymmetric problem: $p, \rho, T, B_{r}, B_{\theta}, v_{r}$, and $v_{\theta}$. In practice, \eqref{eq:gibbs} determines $T$ given $p$ and $\rho$, while \eqref{eq:energy} determines the {the relationship between $p$ and $\rho$, i.e. the barotropic EOS $p = p(\rho)$ initially}. Under the assumption of an ideal gas (consistent with a polytropic EOS), equation \eqref{eq:gibbs} leads to the well-known relation \citep{ST83}
\begin{equation} \label{eq:temp}
T = \frac {\eta m_{u} p} {k_{B} \rho},
\end{equation}
where $\eta$ is the mean molecular weight, $m_{u}$ is the atomic mass unit and $k_{B}$ is the Boltzmann constant. The temperature field $T(p,\rho)$ in \eqref{eq:temp} is calculated from the Grad-Shafranov output and forms an additional input into \PLUTO \,for thermal conduction simulations. We make the assumption of symmetric nuclear matter to determine $\eta$ for simplicity as in previous work [see section 2.3 of \cite{pri11}]. Although the ideal gas law \eqref{eq:temp} is modified in degenerate matter, it provides a good approximation for partially-degenerate, accreted material on a neutron star crust \citep{schatz99} and is straightforward to handle within \PLUTO. A sensitivity analysis associated with expression \eqref{eq:temp} is presented in Appendix B, where it is shown that using \eqref{eq:temp} to determine $T$ as opposed to a degenerate EOS calculated from first principles overestimates the temperature by $\lesssim 15\%$ throughout the bulk of the mountain (see also Secs. 2.3 and 3), where thermal transport matters most; a small effect compared to other uncertainties in the problem.

%The errors introduced as a result of this are small compared to other systematic errors introduced through the various approximations made throughout (see Secs. 2.3 and 3)

\subsection{Thermal conduction}

%For our purposes, equation \eqref{eq:chandra} is to be thought of as an equation governing the evolution of the temperature profile, where $\boldsymbol{F}$ is taken to be a thermal flux whose profile is described below.

In the presence of thermal conduction, the flux on the left-hand side of \eqref{eq:energy} takes the form \citep{LL59}
\begin{equation} \label{eq:heatflux}
\boldsymbol{F} = \kappa_{||} \hat{\bb} ( \hat{\bb} \cdot \bsn T ) + \kappa_{\perp} [ \bsn T - \hat{\bb} ( \hat{\bb} \cdot \bsn T ) ] ,
\end{equation}
where the thermal conductivities $\kappa_{||}$ and $\kappa_{\perp}$, both measured in units of $\textrm{erg} \textrm{ s}^{-1} \textrm{ K}^{-1} \textrm{ cm}^{-1}$, describe heat transport parallel and perpendicular to the magnetic field respectively. The conductivities of a magnetised, fully ionized plasma are dominated by electron transport. They are given in the diffusion approximation by the Balescu-Braginskii formulas [see \cite{braginskii65,pot99} and Table 3.2 of \cite{balescu1}],

\begin{equation}
\kappa_{||} \approx\, 1.77 \times 10^{16} \left( \frac {T_{e}} {10^9 \text{ K}} \right)^{5/2} \textrm{erg} \textrm{ s}^{-1} \textrm{ K}^{-1} \textrm{ cm}^{-1} , \label{eq:estimate1}
\end{equation}
 and
\begin{align} 
%\begin{split}
\kappa_{\perp} \approx&\, \kappa_{||} \Bigg[ 1 + 1.19 \left( \frac {\rho} {10^{10} \textrm{ g cm}^{-3}} \right)^{-2} \left( \frac {|\bb|} {5 \times 10^{13} \text{ G}} \right)^{2} \nonumber  \\
&\times \left( \frac {T_{e}} {10^9 \text{ K}} \right)^{3} \Bigg]^{-1} \textrm{erg} \textrm{ s}^{-1} \textrm{ K}^{-1} \textrm{ cm}^{-1} , \label{eq:estimate2}
%\end{split}
\end{align}
assuming the Coulomb logarithm satisfies $\ln \Lambda = 30 \approx \textrm{constant}$ \citep{bal86}. The estimates \eqref{eq:estimate1} and \eqref{eq:estimate2} are independent of the ideal gas assumption \eqref{eq:temp}. In the limit of a vanishing magnetic field, one has $\kappa_{\perp} = \kappa_{||}$, and \eqref{eq:heatflux} reduces to $\boldsymbol{F} = \kappa_{||} \bsn T$.

%In a typical magnetic mountain environment, \eqref{eq:estimate1} and \eqref{eq:estimate2} imply $\kappa_{\perp} \gtrsim \kappa_{||}$ at the pole $(|\bb| \lesssim 10^{13} \text{ G})$ and $\kappa_{\perp} \lesssim \kappa_{||}$ at the equator $(|\bb| \gtrsim 10^{13} \text{ G})$. %The implication is that hot-spot regions are more likely to form along field lines near the equator but across field lines near the pole. 
%The strong spatial variation of the conductivity coefficients suggests that fenced-off hot-spots are likely to form in regions in between the pole and equator, where one conductivity coefficient begins to dominate over another.

% [Discuss.]

%The `sunspot' model for modifying $T$ in \eqref{eq:temp} by a term proportional to the magnetic pressure $\propto |\bb|^2$ (higher $|\bb|$ implies cooler gas) is considered in Sec. 5 \citep{priest82,pri14}

%The heat flux 

\subsection{Time-scales}
A mountain with the structure in Fig. \ref{mountdiag} contains steep density and magnetic field gradients, so there is no unique definition for characteristic time-scales, like the Alfv{\'e}n time $\tA$ and thermal conduction time $\tC$. In order to analyse our numerical results in Sec. 3 onwards, we adopt the definition \citep{dip12} %[CHECK POWERS AND VALUES]
\begin{align} \label{eq:alfventime}
\tA &= L \sqrt{4 \pi \bar{\rho}}/ |\bar{\bb}| \\
%\begin{split}
&\approx 7 \times 10^{-5} \left( \frac {L} {10^4 \text{ cm}} \right) \left( \frac {\bar{\rho}} {10^{10} \text{ g cm}^{-3}} \right)^{1/2} \nonumber \\
&\,\,\,\,\,\,\times \left( \frac {|\bar{\bb}|} {5 \times 10^{13} \text{ G}} \right)^{-1} \textrm{ s}, \label{eq:alfvenexplicit}
%\end{split}
\end{align}
where $L$ is taken to be the density scale-height where $\rho$ drops to $10^{-3}$ times its maximum value $\rho_{\text{max}}$, $\bar{\rho}$ is the volume-averaged density
\begin{equation} \label{eq:barrho}
\bar{\rho} = \frac{1}{\Vm} \int d^{3} \boldsymbol{x} \, \rho,
\end{equation}
where $\Vm$ is the mountain volume ($ 10^{-7} \rho_{\text{max}} \leq \rho \leq \rho_{\text{max}}$) and $|\bar{\bb}|$ is the volume-averaged magnetic field strength,
\begin{equation} \label{eq:avgb}
|\bar{\bb}| = \frac{1}{\Vm} \int d^{3} \boldsymbol{x} \, |\bb|.
\end{equation}
All the quantities \eqref{eq:alfventime}--\eqref{eq:avgb} are computed at $t=0$ to define $\tA$ for any given run. Similarly, for thermal conduction, from the heat equation we have
\begin{align}
%\begin{split}
\tC =& \frac {5} {3} \left( L / 1 \textrm{cm} \right)^2 \left( \bar{\rho} / \textrm{ g cm}^{-3} \right) \nonumber \\
&\times \left[ \bar{\kappa}_{||}  / \textrm{erg} \textrm{ s}^{-1} \textrm{ K}^{-1} \textrm{ cm}^{-1} \right]^{-1}  \text{ s} \\ 
%\end{split}
&\hspace*{-0.33cm}\approx\ 94.2 \left( \frac {L} {10^4 \textrm{ cm}} \right)^{2} \left( \frac {\bar{\rho}} {10^{10} \textrm{ g cm}^{-3}} \right) \left( \frac {\bar{T}_{e}} {10^9 \text{ K}} \right)^{-5/2}   \textrm{ s} \label{eq:thermaltime},
\end{align}
where $\bar{\kappa}_{\perp}$, $\bar{\kappa}_{||}$, and $\bar{T}_{e}$ are volume-averaged quantities calculated in the same manner as $\bar{\rho}$ in \eqref{eq:barrho} and $|\bar{\bb}|$ in \eqref{eq:avgb}. %Note that we have $\bar{\kappa}_{||} > \bar{\kappa}_{\perp}$. %even though $\kappa_{||} < \kappa_{\perp}$ locally at some points in the mountain.

The thermal conduction time is $\sim 10^{6}$ times longer than the Alfv{\'e}n time for a typical, realistic mountain. Computational expense restricts us to $t \lesssim 2 \tC$ throughout most of this paper (see Sec. 3). The ratio of the time-scales varies from run to run and for different EOS. We perform some `long-term' evolutions (up to $\sim 80 \tC$) in Section 5 to explore the effects of thermal relaxation. In Sections 3 and 4 we show that \eqref{eq:alfvenexplicit} and \eqref{eq:thermaltime} agree with the time-scales of characteristic behaviours observed empirically in the simulations.

%MENTION ZEUS FOR TAUA (maybe in sec. 3.1)

\section{Thermal evolution}

\textsc{\footnotesize{PLUTO}} \citep{pluto} is a general-purpose MHD solver designed to handle steep gradients associated with strong shock phenomena in astrophysical applications. It solves  \eqref{eq:cty}--\eqref{eq:energy}  given the Grad-Shafranov solution and the boundary conditions described in Sec. 2.2 as inputs. The details of the computation are presented in Appendix A along with convergence tests.

In this section we present results from \PLUTO\, simulations of magnetic mountain evolution on time-scales comparable to $\tC$.  We load a Grad-Shafranov equilibrium calculated in Sec. 2.2 for some equation of state (e.g. A,B, or C in Table \ref{tab:table1eos}) into \PLUTO \,and evolve it in two ways, with thermal conduction switched on or off, to explore the mass density and magnetic field profiles for a variety of runs (Sec. 3.2), the evolution of the thermal flux (Sec. 3.3), and the evolution of global observables like $\mu$ (Sec. 4.1) and $\epsilon$ (Sec. 4.2). %We explore.

%In the absence of thermal conduction, we find that the resulting density, pressure, and magnetic field maxima agree with their static Grad-Shafranov counterparts within an accuracy of $\sim 20\%$ after $t \gtrsim 20 \tA$. 

Several numerical and physical issues affect the \PLUTO \,output. (i) The Grad-Shafranov code computes $\psi$, while \PLUTO \,accepts the components of $\boldsymbol{B}$. The calculation of $\boldsymbol{B}$ from $\psi$ involves differentiation, which introduces some numerical error. We use \PLUTO's inbuilt bi-linear interpolation algorithm to map the Grad-Shafranov output to \PLUTO \,input (see Appendix A). (ii) \PLUTO \,maintains ideal-MHD flux freezing through a Godunov scheme [e.g. \cite{godunov05}], but it does not act directly to satisfy the integral constraint \eqref{eq:fluxfreezing} on $M(\psi)$. As $M(\psi)$ is not an input into \PLUTO \,, and equation \eqref{eq:fluxfreezing} is a non-linear equation for $\psi$, it is possible that multiple, valid solutions for $\psi$ exist at any given $t$. One can imagine \PLUTO \,picking a solution branch unpredictably based on numerical fluctuations, if two valid solutions for $\psi$ are numerically close. (iii) The Grad-Shafranov equilibrium may not represent the stable endpoint of a well-posed initial value problem because the Grad-Shafranov equation has multiple unstable solutions \citep{pstab07}. This is related to the loss-of-equilibrium phenomenon investigated by \cite{sturrock89}.

%Since we do not model the Poisson equation $0 = \bsn^2 \phi - 4 \pi G \rho$, we find that the $t=0$ force-balance equation \eqref{eq:somereln} admits the simple solution \eqref{eq:lagcha}. In reality, $\phi$ is also a function of $\rho$, and the Lagrange-Charpit solution \eqref{eq:lagcha} for the function $F$ is more complicated. As such, the $t=0$ force-balance equations are not \emph{exactly} satisfied, which may introduce numerical error when time evolution is applied since \PLUTO \,attempts to build exact, numerical solutions at each timestep. Furthermore, t
%[REF FIGURES -- for THESIS.]

\subsection{Representative example}

We start by considering a representative simulation, which demonstrates the main features of thermal evolution: EOS model A with $M_{a} = 1.8 \times 10^{-5} M_{\odot} \approx 0.58 M_{c}$. We set up a polytropic initial state ($\Gamma = 1.18$ at $t=0$), allow thermal conduction to take place, and evolve the mountain. The EOS parameters, described in Table \ref{tab:table1eos}, are entered into the Grad-Shafranov solver, which produces the initial input for \PLUTO. The thermal profile is entered according to \eqref{eq:temp}. Two separate  \PLUTO \,instances are evolved, with and without the thermal flux $\boldsymbol{F}$ appearing in the right-hand side of equation \eqref{eq:energy}. The time-scales \eqref{eq:alfvenexplicit} and \eqref{eq:thermaltime} read $\tA \approx 2.0 \times 10^{-5} \text{ s}$ and $\tC \approx 5.6 \times 10^{6} \tA$. % Figure \ref{dynamicalrep} presents results from the \PLUTO \,evolution without (left panel) and with (right panel) thermal conduction at $t = 2 \tC$.

%Hence the example respects the the astrophysical condition $\tC \gg \tA$ while keeping $\tC$ short enough to make long-term simulations on the time-scale $\tC$ practical.

\begin{figure*}
\includegraphics[width=\textwidth]{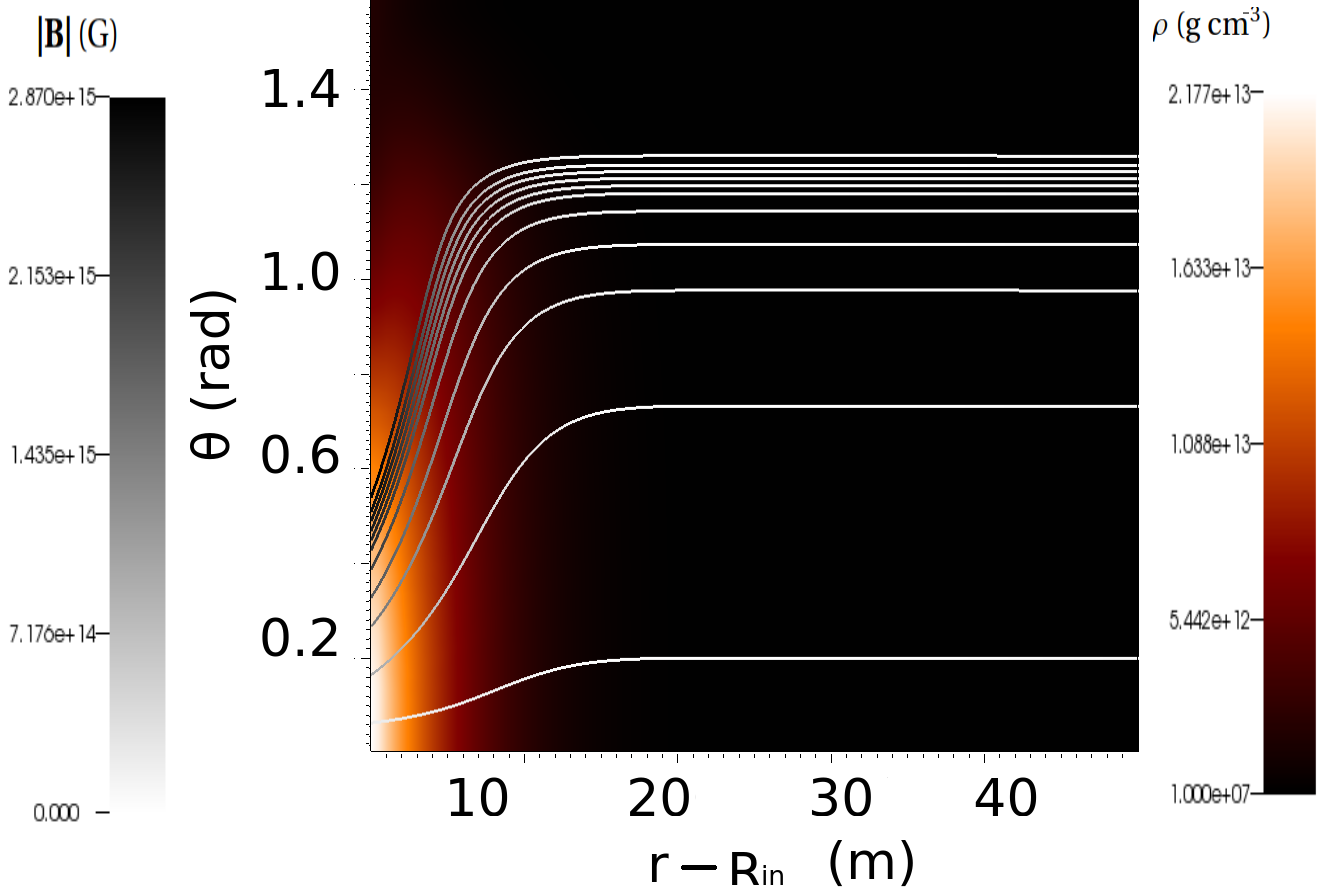}
\caption{Density $\rho$ (colour scale; brighter shades indicate higher $\rho$) and magnetic field lines (darker shades indicate higher $|\bb|$) for the realistic accreted crust model A with accreted mass $M_{a} = 1.8 \times 10^{-5} M_{\odot} \approx 0.58 M_{c}$ at time $t=0$, plotted as functions of altitude (horizontal axis) and colatitude (vertical axis).   \label{rept0}
}
\end{figure*}

In Figure \ref{rept0}, {which demonstrates several features typical of an initially polytropic mountain}, we graph contours of $\rho$ and magnetic field lines at $t=0$. The mountain reaches a maximum altitude of $\Hm \approx 1.3 \times 10^{3} \text{ cm}$ near $\theta = \pi/4$ (where $|\bb|$ rises to a maximum). Most of the mass $(\approx 92 \%)$ is concentrated within the octant $0 \leq \theta \leq \pi/4$. The densest point, with $\rho = 2.2 \times 10^{13} \text{ g cm}^{-3}$, lies at the pole. The magnetic field lines are shifted equatorially; the maximum contour lies at $\theta = \pi/4$, in contrast to the initial dipole field (maximum at $\theta = 0$).

\begin{figure*}
\includegraphics[width=\textwidth]{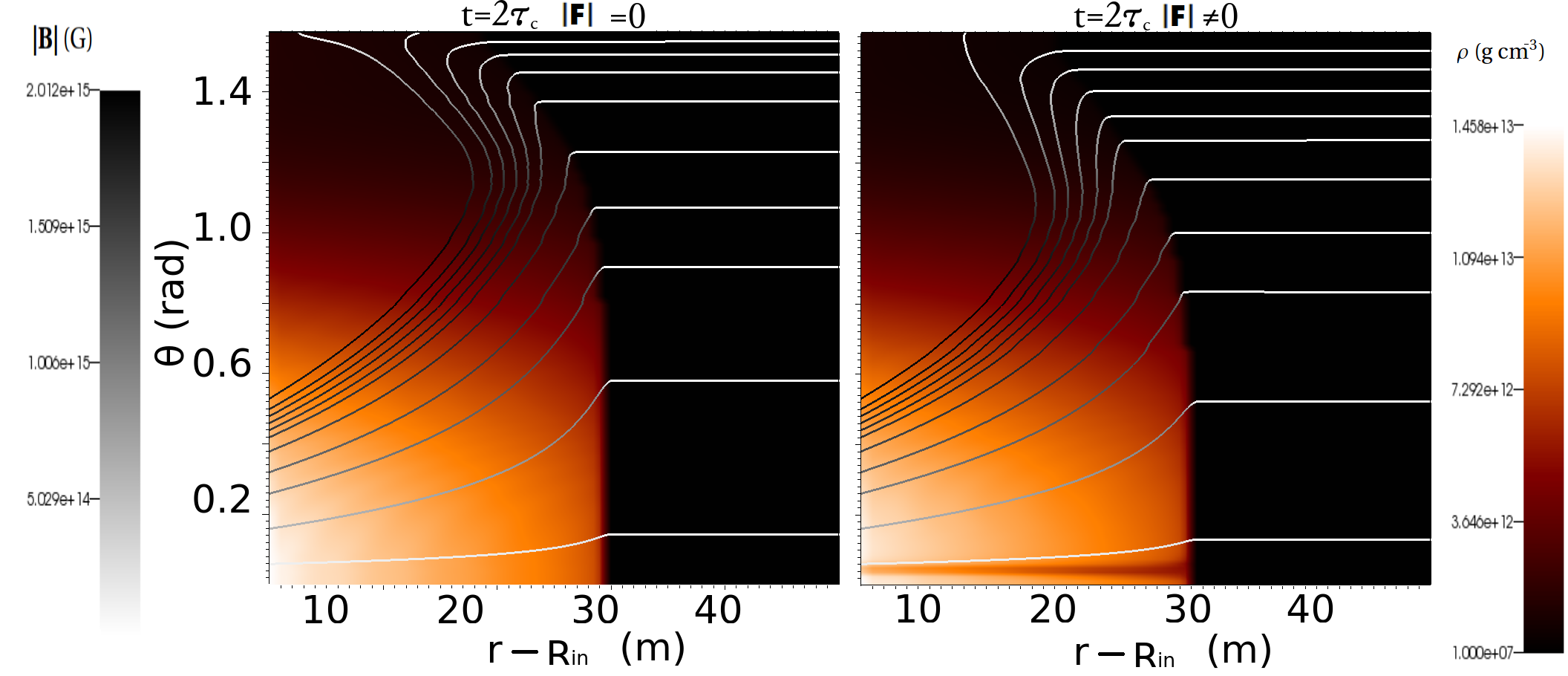}
\caption{Density $\rho$ (colour scale; brighter shades indicate higher $\rho$) and magnetic field lines (darker shades indicate higher $|\bb|$) for the realistic accreted crust model A with accreted mass $M_{a} = 1.8 \times 10^{-5} M_{\odot} \approx 0.58 M_{c}$ with heat flux $\boldsymbol{F} =0$ (left panel) and $\boldsymbol{F} \neq 0$ (right panel), plotted as functions of altitude (horizontal axis) and colatitude (vertical axis). The snapshots are taken at time $t=2\tC$. \label{dynamicalrep}
}
\end{figure*}

Figure \ref{dynamicalrep} {presents results from \PLUTO \,} and illustrates how thermal conduction affects the evolution. It shows snapshots at $t=2 \tC$ of an adiabatic mountain ($\boldsymbol{F} = 0$, left panel) and one evolved with a nonzero thermal flux ($\boldsymbol{F} \neq 0$, right panel). The mountain grows from $\Hm \approx 1.3 \times 10^{3} \text{ cm}$ to $\Hm \approx 3.0 \times 10^{3} \text{ cm}$. It is $\approx 4\%$ taller for the run with conduction, but its density is lower (by a factor $\lesssim 5$ near $\theta = 0.1$ where an underdense column forms; {see Sec. 3.2}) everywhere except at $\theta \approx 0$ and $\theta \approx 0.2$. {This is a consequence of the continuity equation \eqref{eq:cty}, which demands that an increase in height is met with an overall decrease in mass density.} Aside from the underdense column, thermal conduction has the effect of driving matter towards the pole, where the density attains a maximum of $\rho_{\text{max}} = 1.3 \times 10^{13} \text{ g cm}^{-3}$ for the run without conduction (left panel) and $\rho_{\text{max}} = 1.5 \times 10^{13} \text{ g cm}^{-3}$ for the run with conduction (right panel) ($\approx 16 \%$ increase).   An analogous thermal softening phenomenon occurs in crustquake models, where thermal transport amplifies shear stresses felt in the neutron star crust \citep{chug10,belo14}. 

%We observe that the additional compression of matter at the pole run with thermal conduction does not lead to a similar increase in the pressure, thereby suggesting a `softening' of the effective EOS.

Evolution with $\boldsymbol{F} \neq 0$ tends to widen the magnetic field contours (cf. Fig. \ref{rept0}), because the mountain spreads and drags the field-lines with it [the time-dependent version of the flux-freezing condition \eqref{eq:fluxfreezing}]. {Thermal conduction causes matter to be shifted both towards the pole and towards the base of the mountain, causing magnetic `pockets' to form near $\theta = 0$ {(see also Sec. 3.2 and Fig. 6, where they are clearer)}, as the field lines bend around the drifting matter.} Overall, the magnetic field is weakened, going from a maximum strength of $|\bb|_{\text{max}} = 2.9 \times 10^{15} \text{ G}$ to $|\bb|_{\text{max}} = 1.9 \times 10^{15} \text{ G}$ and $|\bb|_{\text{max}} = 2.0 \times 10^{15} \text{ G}$ at $t= 2 \tC$ without and with conduction, respectively. {Away from the pole, the locations of the maxima and minima of $|\boldsymbol{B}|$ are largely unaffected by conduction.}
%, and the field strength differs by a maximum of $6\%$ at $\theta = \pi /4$. 

%\begin{figure}
%\includegraphics[width=0.473\textwidth]{TCfinal2.png}

%FIG. 4 -- Density and magnetic field contours for an adiabatic EOS (model B) with accreted mass $M_{a} = 2 \times 10^{-8} M_{\odot} \approx 0.38 M_{c}$ at time $t=25 \tA$ with heat flux $\boldsymbol{F}$ given by \eqref{eq:heatflux} in terms of the altitude (horizontal axis) and latitude (vertical axis). The density is shown on the continuous colour spectrum, with brighter shades indicating a greater density. The magnetic field is represented by the solid contours, with darker shades indicating a greater field strength. 
%\end{figure}

\subsection{Mass density and magnetic field evolution}

%In Figure 2 we present contour plots for mountain density profiles at $t=0$ for a representatives model A with $M_{a} =  5 \times 10^{-5} M_{\odot}$ (left panel), B with $M_{a} = 2 \times 10^{-8} M_{\odot}$ (middle panel), and C with $M_{a} = 10^{-6} M_{\odot}$ (right panel). In Figures 3 and 4 we present contour plots for the mountain density profile at late times $t \gtrsim 20 \tA$ having evolved without (Figure 3) and with (Figure 4) thermal conduction, where the mountain parameters match those as in Figure 2. 

\begin{figure*}
\includegraphics[width=\textwidth,height=0.92\textheight]{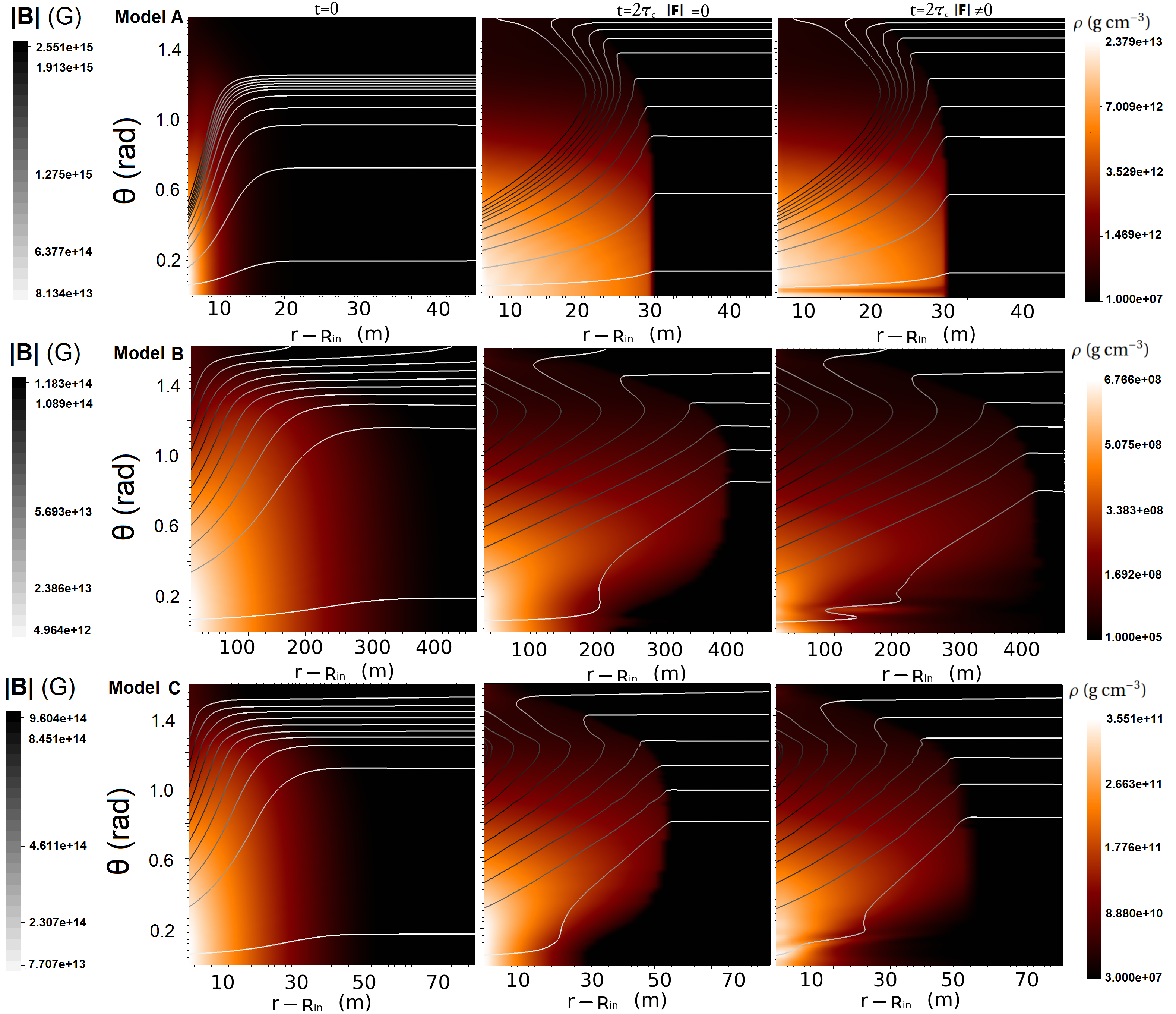}
\caption{Density $\rho$ (colour scale; brighter shades indicate higher $\rho$) and magnetic field lines (darker shades indicate higher $|\bb|$) for a {realistic accreted crust} EOS (model A, top row; $M_{a} = 2.4 \times 10^{-5} M_{\odot} \approx 1.0 M_{c}$), and two {idealised polytropic} EOS (model B, middle row; $M_{a} = 3.0 \times 10^{-8} M_{\odot} \approx 1.2 M_{c}$; model C, bottom row; $M_{a} = 2.0 \times 10^{-6} M_{\odot} \approx 1.0 M_{c}$) studied by Priymak et al. (2011), at times $t=0$ (left panel) and $t=2\tC$ with $\boldsymbol{F} =0$ (middle panel) and $t=2\tC$ with $\boldsymbol{F} \neq 0$ (right panel). The fields are plotted as functions of altitude (horizontal axis) and colatitude (vertical axis).  \label{eosdenmag}
}
\end{figure*}

In Figure \ref{eosdenmag} we plot $\rho$ contours and {magnetic field lines} for model A (top row) with $M_{a} =  2.4 \times 10^{-5} M_{\odot} \approx 1.0 M_{c}$, model B (middle row) with $M_{a} = 3.0 \times 10^{-8} M_{\odot} \approx 1.2 M_{c}$, and model C (bottom row) with $M_{a} = 2.0 \times 10^{-6} M_{\odot} \approx 1.0 M_{c}$ for times $t=0$ (left panel) and $t = 2 \tC$ without (middle panel) and with (right panel) thermal conduction. The initial state is read from the Grad-Shafranov output and is the same for both runs for any given EOS.  All three runs have similar\footnote{Ideally we would keep $M_{a}/M_{c}$ exactly the same across all runs. However, this is impractical because one does not know what $\mu$ is \emph{prior} to running the Grad-Shafranov code for a given $M_{a}$, and it is computationally expensive to try and tune $M_{a}/M_{c}$ exactly.} values of $M_{a}/M_{c} \sim 1$, implying that $\mu$ at $t=0$ is approximately halved by burial in all three models (see Sec. 4.1). {We emphasise that models B and C are poor approximations to a realistic crust. They are included throughout mainly to give the reader a general sense of how the mountain structure depends on the EOS as well as to make contact with previous work for completeness.}

%Figures 6 and 7 are laid out identically to Fig. 5 for model B (Fig. 6) with accreted mass $M_{a} = 3 \times 10^{-8} M_{\odot} \approx 1.18 M_{c}$ and model C (Fig. 7) with accreted mass $M_{a} = 2 \times 10^{-6} M_{\odot} \approx 1.02 M_{c}$.

{
The model A mountain (top row of Fig. \ref{eosdenmag}) grows taller over time, thereby reducing $\bar{\rho}$; we find $\Hm = 3.3 \times 10^{3} \text{ cm}$ for both $\boldsymbol{F} = 0$ and $\boldsymbol{F} \neq 0$ at $t= 2\tC$, compared to $\Hm = 1.2 \times 10^{3} \text{ cm}$ at $t=0$. The magnetic pole remains the densest region with maximum densities $\rho_{\text{max}} = 2.38 \times 10^{13} \text{ g cm}^{-3}$ at $t=0$ and $\rho_{\text{max}} = 1.43 \times 10^{13} \text{ g cm}^{-3}$ at $t= 2 \tC$ without thermal conduction and $\rho_{\text{max}} = 1.45 \times 10^{13} \text{ g cm}^{-3}$ at $t= 2 \tC$ with thermal conduction. A narrow, underdense $(\rho \lesssim 4 \times 10^{12} \text{ g cm}^{-3}$) column forms at $\theta \approx 0.1$ for the run with thermal conduction. Field lines are noticeably distorted from their $t=0$ state by flux freezing through a combination of poleward flow, which pushes them towards $\theta = 0$, and stretching caused by the $\sim$ three-fold increase in $\Hm$, which pushes them radially outward.}

{Equation \eqref{eq:temp} implies $\bsn T \propto \bsn \rho^{\Gamma - 1}$. Hence we expect thermal conduction to be less influential in model A $(\Gamma = 1.18)$ than in models B and C $(\Gamma = 5/3)$.} The evolution of adiabatic model B (middle row of Fig. \ref{eosdenmag}) is noticeably affected by the non-zero flux term $\boldsymbol{F}$ (cf. middle and right panels). The matter column near the pole has height $\Hm(\theta \approx 0) = 1.8 \times 10^{4} \text{ cm}$ for $\boldsymbol{F} = 0$ and $\Hm(\theta \approx 0) = 3.0 \times 10^{4} \text{ cm}$ for $\boldsymbol{F} \neq 0$. Matter concentrates more at the base of the mountain $(r \approx R_{\text{in}})$ for $\boldsymbol{F} \neq 0$, reaching peak densities of $\rho_{\text{max}} = 6.8 \times 10^{8} \text{ g cm}^{-3}$ for $\boldsymbol{F} = 0$ and $\rho_{\text{max}} = 9.2 \times 10^{8} \text{ g cm}^{-3}$ for $\boldsymbol{F} \neq 0$ at $\theta = 0$. Both evolved mountains are denser than the initial state $(\rho_{\text{max}} = 6.4 \times 10^{8} \text{ g cm}^{-3})$. Thermal conduction drives matter towards the pole, like what is seen in Fig. \ref{dynamicalrep}. The mountain grows taller, albeit comparatively less so than for model A, going from peak altitudes $\Hm \approx 4.0 \times 10^{4} \text{ cm}$ to $\Hm \approx 4.2 \times 10^{4} \text{ cm}$ with $\boldsymbol{F} = 0$ and $\Hm \approx 4.3 \times 10^{4} \text { cm}$ for $\boldsymbol{F} \neq 0$.

%The mountain density is equatorially shifted as time evolves, and develops a hill-like structure centred about $\theta = \pi/4$ similar to model A except less prominently $(\rho \sim 5 \times 10^{7} \text{ g cm}^{-3} \sim 10^{1} \rho_{\text{max}})$. 

%As for the isothermal case, the magnetic pole remains the densest region, with densities reaching $\rho \sim 8 \times 10^{8} \text{ g cm}^{-3}$ there. 

Adiabatic model C (bottom row of Fig. \ref{eosdenmag}) evolves like model B. The mountain grows taller on the conduction time-scale, going from $\Hm \approx 4.0 \times 10^{3} \text{ cm}$ at $t=0$ to $\Hm \approx 5.0 \times 10^{3} \text{ cm}$ for $\boldsymbol{F} = 0$ and $\Hm \approx 5.3 \times 10^{3} \text { cm}$ for $\boldsymbol{F} \neq 0$. At $t=0$, the density maximum $\rho_{\text{max}} = 3.7 \times 10^{11} \text{ g cm}^{-3}$ lies at $r \approx R_{\text{in}}$ and $\theta=0$. After evolution the density reaches maximum values at the same location of $\rho_{\text{max}} = 3.5 \times 10^{11} \text{ g cm}^{-3}$ ($\approx 5\%$ decrease) for $\boldsymbol{F} = 0$ and $\rho_{\text{max}} = 4.0 \times 10^{11} \text{ g cm}^{-3}$ ($\approx 8\%$ increase) for $\boldsymbol{F} \neq 0$  at $t=2\tC$. The compression of matter at the pole suggests that $\boldsymbol{F} \neq 0$ acts to `soften' the effective EOS.

%Mass conservation, as before, then implies that the rest of the mountain is comparatively sparse. 

%We see a narrow column develop between $1.45 \leq \theta \leq \pi/2$ with strong densities $\rho \sim 5 \times 10^{10} \text{ g cm}^{-3} \sim 10^{-1} \rho_{\text{max}}$ throughout, reaching an altitude of $\Hm \sim 55 \text{ m}$. This region is absent in the $t=25\tA$ state of the no-conduction run.

%\subsection{Magnetic field}

%We investigate the magnetic field structure.

%Figures 8 through 10 display magnetic field strength extracted from the runs performed Figures 5 through 7. We see that

By inspecting the magnetic field lines in Fig. \ref{eosdenmag}, we see that $|\boldsymbol{B}|$ evolves similarly to $\rho$ due to flux freezing. As discussed above, thermal conduction drives matter towards the pole, shifting $\boldsymbol{B}$ accordingly \citep{pm04,pri11}. Hence $|\bb|$ decreases on the whole as time passes, most dramatically in the case of model C, which predicts $|\bb|_{\text{max}} = 9.7 \times 10^{14} \text{ G}$ at $t=0$, $|\bb|_{\text{max}} = 6.5 \times 10^{14} \text{ G}$ for $\boldsymbol{F} = 0$ at $t=2\tC$, and $|\bb|_{\text{max}} = 6.6 \times 10^{14} \text{ G}$ for $\boldsymbol{F} \neq 0$. This is similar to what occurs for the representative example discussed in Sec. 3.1 and tests with different grid resolutions (not plotted), {suggesting the possibility that} $|\boldsymbol{B}|_{\text{max}}$ increases slightly for runs with conduction, independent of the EOS. {Additional, higher-resolution convergence tests (cf. Appendix A) can be undertaken, if future observational applications warrant.} The formation of dense filamentary regions $(\rho \sim 10^{-1} \rho_{\text{max}})$ for runs with $\boldsymbol{F} \neq 0$ at $\theta \approx 1.0$ causes several magnetic `pockets' to form near the pole, as the magnetic field lines bend around poleward-drifting matter. {The formation of filaments near the magnetic pole, as observed across all simulations with thermal conduction, may stem from thermal Parker-like instabilities \citep{park1953,field1965}. These instabilities introduce `finger-like' density structures, which emerge due to the propagation of contact discontinuities between lighter and denser sections of fluid \citep{stone07,mous09}. Although we only have one fluid in our model, the strong dependence of the conduction coefficient \eqref{eq:estimate2} on the local magnetic field strength, which varies strongly near the pole, may cause this `fenced-off' behaviour. It has been shown that unstable modes grow faster in the presence of anisotropic thermal conduction \citep{leco12}.}

\subsection{Heat flux}

%In Figure \ref{eosdenmag} we plot $\rho$ and $|\bb|$ contours for model A (top row) with $M_{a} =  5 \times 10^{-4} M_{\odot} \approx 0.98 M_{c}$, model B (middle row) with $M_{a} = 3 \times 10^{-8} M_{\odot} \approx 1.18 M_{c}$, and model C (bottom row) with $M_{a} = 2 \times 10^{-6} M_{\odot} \approx 1.02 M_{c}$ for times $t=0$ (left panel) and $t = 2 \tC$ without (middle panel) and with (right panel) thermal conduction.

\begin{figure*}
\includegraphics[width=\textwidth,height=0.92\textheight]{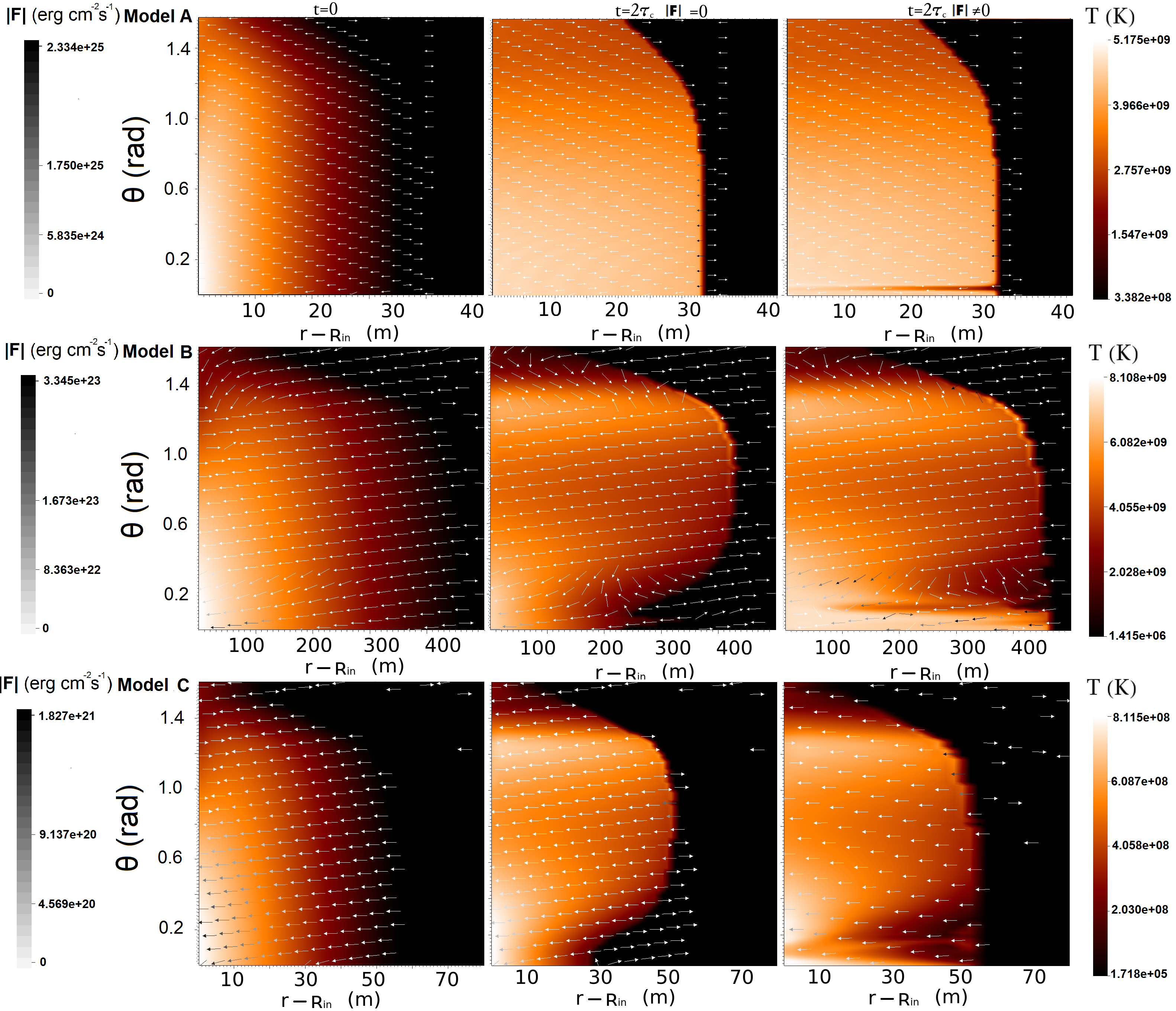}
\caption{Temperature $T$ (colour scale; brighter shades indicate higher $T$) and thermal flux $\boldsymbol{F}$ (vectors; darker arrows indicate higher $|\boldsymbol{F}|$). The layout and simulation parameters are the same as in Fig. \ref{eosdenmag}.  {Although the simulations in the middle column are run with $\boldsymbol{F} = 0$, we can still compare what a theoretical $\boldsymbol{F}$ might be from \eqref{eq:heatflux}, if conduction is switched on at some time $t > 0$.} \label{eostempflux}
}
\end{figure*}

Figure \ref{eostempflux} displays contours of temperature $T$ (colour scale) and the direction of the thermal flux $\boldsymbol{F}$ (arrows) extracted from the runs performed in Figure \ref{eosdenmag}. We seek to identify the existence, and evolution, of thermal hot spots [e.g. \cite{thermfluxpaper}]. %Regions where $|\boldsymbol{F}|$ is large indicate that the divergence term $\bsn \cdot \boldsymbol{F}$ within \eqref{eq:energy} is strong, and as such there is a dominating heat source in those regions.

{
The temperature varies gradually with $r$ and $\theta$ for model A (top row), because the polytropic index $\Gamma$ is nearly unity. Nevertheless, the maximum (at the pole) and minimum (at the mountain-atmosphere interface) values of $T$ are in the ratio $T_{\text{max}}/T_{\text{min}} \approx 10$, a significant contrast. The thermal flux is predominantly directed towards the base of the mountain at all times, but becomes more `noisy' at large $t$, when local hot spots form. At $t=2\tC$ we see that the temperature profile becomes more uniform away from the pole, suggesting that the model evolves towards an isothermal end state [$d/dt\left(p/\rho\right) \approx 0$ for $t \gg \tC$], even when thermal conduction is not implemented.  At $\theta \approx 0.1$ for the run with $\boldsymbol{F} \neq 0$ (right panel) we see a region of relatively low temperature $(T_{\text{max}}/T \sim 6)$ form. This `heat sink' is underdense as seen in Fig. \ref{eosdenmag}, and is surrounded by the local hot spots $(T_{\text{spots}} \sim 5 \times 10^{9} \text{ K})$ described above.} %We note that  $\bsn T$ is smaller by $\sim 5$ per-cent everywhere outside of this sink at $t= 2\tC$ for $\boldsymbol{F} \neq 0$ as compared with the $\boldsymbol{F} = 0$ case, further suggesting that thermal conduction runs are more isothermal at late times $t \gtrsim 2 \tC$.

For Model B (middle row) we have $|\boldsymbol{F}| \approx 10^{23} \text{ erg} \text{ cm}^{-2} \text{s}^{-1}$, and $\boldsymbol{F}$ is predominantly directed towards the pole at $\theta =0$. In the run without conduction (middle panel) we see that $\boldsymbol{F}$ is almost indistinguishable from its $t=0$ counterpart except near the equator where heat flows into a hot column $(T \approx 6 \times 10^{9} \text{ K})$ at $\theta = 1.3$. The temperature evolves like the density, i.e. growing and spreading with $\rho$ (cf. Fig. \ref{eosdenmag}). When conduction is switched on, a hot region ($T \gtrsim 8 \times 10^{9} \text{ K}$) forms near the pole which extends to the mountain-atmosphere interface at $\Hm \approx 4.2 \times 10^{4} \text{ cm}$. The flux is highest at the mountain-atmosphere interface and at altitude $r-R_{\text{in}} \approx 1.3 \times 10^{4} \text{ cm}$. The flux is directed in different directions throughout the column, suggesting that localised hot spots form in the densest part of the mountain. Away from the pole $(\theta \gtrsim 0.1)$, the heat flow is small $(|\boldsymbol{F}| \lesssim 10^{-2} |\boldsymbol{F}|_{\text{max}})$.

%It is directed towards the base of the mountain at $\theta=0$, suggesting that $\boldsymbol{F}$ acts to homogenize $T$ in regions where $\rho$ is large.

In Model C (right panel) at $t=0$, we see that heat flows towards the pole $(|\boldsymbol{F}| \approx 10^{21} \text{ erg} \text{ cm}^{-2} \text{s}^{-1})$ and away from the equator $(|\boldsymbol{F}| \approx 10^{20} \text{ erg} \text{ cm}^{-2} \text{s}^{-1})$. At $t=2\tC$, however, heat flows from the top of the mountain to the base near the equator $(\theta \approx 1.3)$, and little heat ($|\boldsymbol{F}| \lesssim 10^{-4} |\boldsymbol{F}|_{\text{max}}$) flows near the pole, where a hot column $(T \approx 8 \times 10^{8} \text{ K})$ develops in a manner to similar to model B.  Because of relation \eqref{eq:temp}, the temperature profile evolves like the density and increases as the mountain grows $(T \propto \rho^{\Gamma-1}$); cf. Fig \ref{eosdenmag}. Overall, the {initially polytropic} mountains respond similarly to thermal conduction by forming hot spots near the equator at $\theta \approx 1.3$ (models B and C) and near the pole at $\theta \approx 0.1$ (all models), where $|\boldsymbol{F}|$ is largest.

\section{Global observables}

In this section we consider the evolution on the conduction time-scale $\tC \gg \tA$ \eqref{eq:thermaltime} of the global observables $\mu$ (Sec. 4.1) and $\epsilon$ (Sec. 4.2) derived from runs of models A, B, and C. Simulations of mountains on the Alfv{\'e}n time-scale $\tA$ have been performed previously using the codes \ZEUS \,\citep{pstab07} and \PLUTO \,\citep{dip13a,dip13b}.

 %In Section 4.3 we consider the evolution of these observables for a particular case (model B) on the conduction time-scale \eqref{eq:thermaltime}, to explore the effects of long-term thermal relaxation on the mountain structure.

\subsection{Magnetic dipole moment}

The theory of magnetic burial predicts that the global magnetic dipole moment for an axisymmetric mountain,
\begin{equation} \label{eq:dipmom}
\mu = \frac {3 r^4} {4} \int^{1}_{-1} d \left( \cos \theta \right) \cos\theta B_{r}(r,\theta),
\end{equation}
evaluated at $r \gtrsim R_{\text{m}}$, decreases as a function of $M_{a}$ \citep{bil98,melphi01}. In order to explore the relationship between burial, accreted mass, and thermal conduction for different EOS, we calculate $\mu$ from for a variety of \PLUTO \,simulations with thermal conduction switched on.

%We confirm the trend for models A, B, and C with and without conduction \citep{bil98,pm04,mp05,vigstab,pri11}. 

\begin{figure*}
\includegraphics[width=\textwidth]{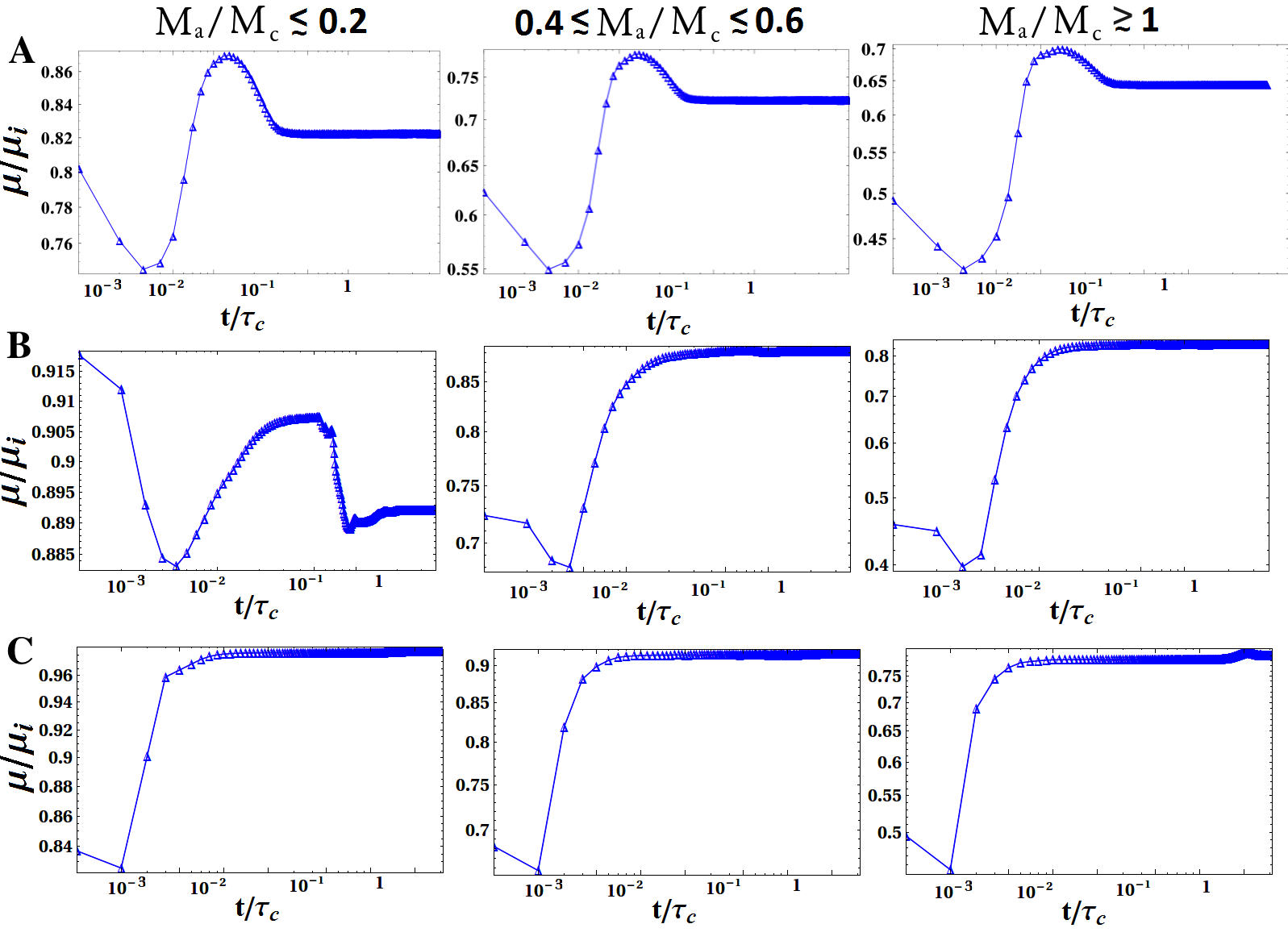}
\caption{Evolution due to thermal conduction of the magnetic dipole moment $\mu$ [equation \eqref{eq:dipmom} evaluated at $r = R_{\textrm{m}}$] normalised to the pre-accretion value $\mu_{i}$ for model A (top row) with accreted masses $M_{a} = 1.2 \times 10^{-5} M_{\odot} \approx 0.23 M_{c}$ (left panel), $M_{a} = 1.8 \times 10^{-5} M_{\odot} \approx 0.58 M_{c}$ (middle panel), and $M_{a} = 2.4 \times 10^{-5} M_{\odot} \approx 1.02 M_{c}$ (right panel), model B (middle row) with $M_{a} = 10^{-8} M_{\odot} \approx 0.09 M_{c}$ (left panel), $M_{a} = 2.0 \times 10^{-8} M_{\odot} \approx 0.38 M_{c}$ (middle panel), and $M_{a} = 3.0 \times 10^{-8} M_{\odot} \approx 1.18 M_{c}$ (right panel), and model C with $M_{a} = 1.0 \times 10^{-6} M_{\odot} \approx 0.19 M_{c}$ (left panel), $M_{a} = 1.5 \times 10^{-6} M_{\odot} \approx 0.47 M_{c}$ (middle panel), and $M_{a} = 2.0 \times 10^{-6} M_{\odot} \approx 1.02 M_{c}$ (right panel). Time is plotted in units of $\tC$ on a logarithmic scale to capture both the MHD $(t \ll \tC)$ and thermal $(t \gtrsim \tC)$ evolution.
\label{eosdipmom}
}
\end{figure*}

Figure \ref{eosdipmom} shows how $\mu$ (normalized to the pre-accretion value $\mu_{i}$) evolves due to thermal conduction for various accreted masses ($0.1 \leq M_{a}/M_{c} \leq 1$; left to right) and {initially adiabatic} EOS ($1.18 \leq \Gamma \leq 5/3$; top to bottom). Each panel displays $\mu(t)$ on a logarithmic temporal scale to capture both the MHD $(t \ll \tC)$ and thermal $(t \gtrsim \tC)$ dynamics. {Again we emphasise that model A (top row of Fig. \ref{eosdipmom}) corresponds most closely to an astrophysically realistic accreted crust. Models B and C are included for completeness to illustrate EOS-related trends and make contact with previous work \citep{pri11}.}

%in physical units of time for isothermal mountains (model A, top row) with $M_{a} = 1.5 \times 10^{-4} M_{\odot} \approx 0.15 M_{c}$ (left panel), $M_{a} = 3.4 \times 10^{-4} M_{\odot} \approx 0.5 M_{c}$ (middle panel), and $M_{a} = 5 \times 10^{-4} M_{\odot} \approx 0.98 M_{c}$ (right panel),  for adiabatic models B (middle row) with $M_{a} = 10^{-8} M_{\odot} \approx 0.09 M_{c}$ (left panel), $M_{a} = 2 \times 10^{-8} M_{\odot} \approx 0.38 M_{c}$ (middle panel), and $M_{a} = 3 \times 10^{-8} M_{\odot} \approx 1.18 M_{c}$ (right panel), and C (bottom row) with $M_{a} = 10^{-6} M_{\odot} \approx 0.19 M_{c}$ (left panel), $M_{a} = 1.5 \times 10^{-6} M_{\odot} \approx 0.47 M_{c}$ (middle panel), and $M_{a} = 2 \times 10^{-6} M_{\odot} \approx 1.02 M_{c}$ (right panel), evolved with thermal conduction. 

All the mountains depicted in Fig. \ref{eosdipmom} undergo an initially violent phase within $\lesssim 10^{2} \tA$, during which $\mu$ drops then rises. The behaviour observed in model A is similar to what is seen in Figures 6 and 14 of \cite{vigstab} for example. It is largely driven by the MHD reconfiguration of the mountain rather than thermal conduction $(10^{2} \tA \ll \tC)$  \citep{vigstab,dip13a}. We find that $\mu$ decreases slightly for all mountains (maximum of $\approx 7\%$ for model A with $M_{a}/M_{c} \approx 0.2$), independent of the EOS, from $t=0$ to $t \sim 10^{2} \tA$, consistent with previous \ZEUS \,simulations \citep{pstab07}. Note that the Grad-Shafranov equilibria, and hence the evolution, are insensitive to the exact value of the initial dipole moment $\mu_{i}$ {provided that we have $M_{a} / M_{c} \lesssim 10$ [cf. the scaling law (1) introduced by \cite{shibaz}] \citep{pm04,pstab07}. In this context, insensitive means that $\mu/\mu_{i}$ depends on $\mu_{i}$ only through the ratio $M_{a}/M_{c}$ and not on $\mu_{i}$ in isolation. Since \cite{pri11} found that $M_{c} \propto \mu_{i}^2$, the insensitivity condition $M_{a} / M_{c} \lesssim 10$ translates into an EOS-dependent lower bound for $\mu_{i}$. For the astrophysically relevant model A, we require [see expression (B26) of \cite{pri11}]
\begin{equation}
\left( \frac {\mu_{i}} {3.2 \times 10^{30} \text { G cm}^{3}} \right)^2 \gtrsim 0.4 \left( \frac {M_{a}} {10^{-4} M_{\odot}} \right) \left( \frac {R_{\star}} {10^{6} \text{ cm}} \right)^{6},
\end{equation}
which is safely applicable to many LMXB systems, at least within the early stages of accretion \citep{hue95,zhang06}. For $M_{a} / M_{c} \gtrsim 10$, the Grad-Shafranov modelling breaks down, and it is an open question whether the results are sensitive to $\mu_{i}$ or not.}

On the longer time-scale $t \gtrsim \tC$, the behaviour of $\mu$ is {qualitatively similar for all three initially polytropic EOS}. In all cases, $\mu$ increases beyond $\mu_{i}$; runs with $M_{a}/M_{c} \gtrsim 0.1$ lead to $\mu(t \gtrsim \tC) > \mu(t=0)$. In effect thermal conduction resurrects some of the buried field, e.g. $\mu$ increases by $\approx 14\%$ in the case of model C with $M_{a}/M_{c} \approx 0.2$. For runs with $M_{a}/M_{c} \gtrsim 0.5$, $\mu$ increases significantly from its initial value at $t=0$, e.g. by up to $\approx 80\%$ in the case of model B for $M_{a} = 3.0 \times 10^{-8} M_{\odot} \approx 1.2 M_{c}$.

{For a realistic accreted crust (model A), we find $0.64 \leq \mu/\mu_{i} \leq 0.83$ for $0.2 \leq M_{a}/M_{c} \leq 1$ at $t = 2 \tC$. Small changes in $\mu$ for $t \gtrsim \tC$ in model A suggest that conduction plays a comparatively minor role in the evolution of astrophysically realistic mountains. Nevertheless we find that substantial ($\mu/\mu_{i} \lesssim 0.5$) magnetic burial requires significantly greater accreted masses than previously estimated by \cite{pri11}; for example, $\mu(t=2\tC)/\mu(t=0) = 1.28$ for $M_{a}/M_{c}(t=0) \approx 1$ suggests an increase in the characteristic mass $M_{c}$ at $t = 2 \tC$ by a factor $\sim 2$. }

{The inclusion of thermal conduction has the effect of partially resurrecting the buried field by increasing $\mu$, which is similar to `softening' the EOS [as found by \cite{pri11}].} The comparatively small increase in $\mu$ for model A (see Fig. \ref{eosdipmom}) implies that the realistic EOS softens less than for models B and C. This is expected because the polytropic index $\Gamma  =1.18$ is closer to unity (i.e. nearly isothermal), implying that $\bsn T$ is smaller than for the isentropic gas models B and C.

%Overall we see that the effects of thermal conduction are minimal on time-scales $t \sim 25 \tA$, especially with respect to the isothermal case, as expected. The curves for $\mu$ are indistinguishable for model A with $M_{a} = 5 \times 10^{-4} M_{\odot}$. We see a very small deviation $(\sim 0.2 \%)$ after $t = 25 \tA$ for the case $M_{a} = 3.4 \times 10^{-5} M_{\odot}$, and a $\sim 1 \%$ discrepancy can be seen in $\mu$ for $M_{a} = 1.5 \times 10^{-4} M_{\odot}$. The effects are more pronounced for larger accreted masses in general for models B and C, especially in model B where we we see a discrepancy of $\sim 5 \%$ between the $t = 25 \tA$ states with and without thermal conduction (middle and right panels).

\subsection{Mass ellipticity}
The characteristic gravitational wave strain emitted by a continuous-wave source is \citep{thorne80,brady}
\begin{equation} \label{eq:wavestrain}
h_{c} = \left(\frac {128 \pi^4} {15}\right)^{1/2} \frac {G I_{zz} \nu^2 |\epsilon|} {d c^2}, 
\end{equation}
\\
where $I_{jk}$ is the moment-of-inertia tensor, $\nu$ is the spin frequency, $d$ is the distance from the Earth to the source, and $\epsilon$ is the mass ellipticity,
\begin{equation} \label{eq:epsilon}
\epsilon = \frac {I_{xx} - I_{yy}} {I_{zz}}.
\end{equation}
The magnitude of $\epsilon$ represents the primary uncertainty in estimating $h_{c}$ in practical astrophysics applications [see e.g. \cite{aasi14,MSM15,geppert16}; though cf. \cite{suvorov18}]. Here we can calculate $\epsilon$ directly from \eqref{eq:epsilon} using $\rho$ as output by \PLUTO. Thus we can explore the effects of thermal conduction on the detectability of magnetic mountains using ground-based interferometers such as the Laser Interferometer Gravitational-Wave Observatory (LIGO) \citep{abb08,hask15}.

\begin{figure*}
\includegraphics[width=\textwidth]{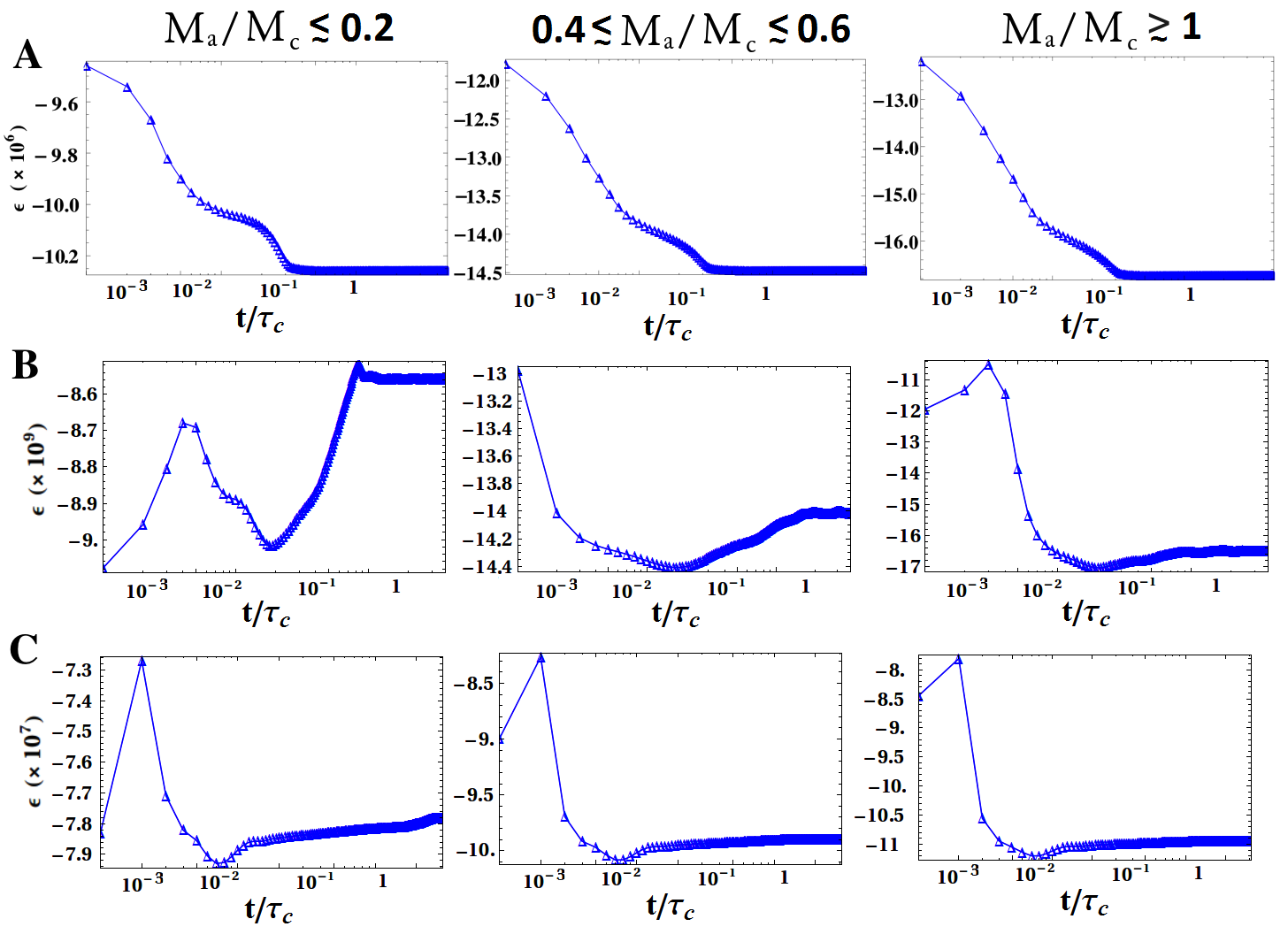}
\caption{Evolution of the mass ellipticity $\epsilon$ evaluated from \eqref{eq:epsilon}. The layout and simulation parameters are the same as in Fig. \ref{eosdipmom}.
\label{eosepsilon}
}
\end{figure*}

Figure \ref{eosepsilon} plots $\epsilon$ against time (in units of $\tC$) for different EOS and values of $M_{a}$. The layout is the same as in Fig. \ref{eosdipmom}. All of the runs yield $\epsilon < 0$, indicating that the star is prolate\footnote{{Since \cite{vigstab} found that three-dimensional simulations of magnetic mountains relax to an almost axisymmetric state after a few Alfv{\'e}n times (see Footnote 1), we expect the star to be prolate even without the assumption of axial symmetry.}}; the mountain is densest at the magnetic pole \citep{cutler02,mmra11}. 

%For model A, $|\epsilon|$ decreases on the time-scale $t \gtrsim \tC$ by up to a factor $\approx 2$ for $M_{a}/M_{c} \approx 1$. This is likely due to the formation of the density column near $\theta = 1.2$ (cf. top row of Fig. \ref{eosdenmag}), which gives the star a more oblate shape and reduces $|\epsilon|$ overall. 

In contrast to the {magnetic dipole moment (Sec. 4.1)}, $|\epsilon|$ increases with time for $t \gtrsim 10^{2} \tA$ for all runs with $M_{a}/M_{c} \gtrsim 0.1$, by up to $\approx 45 \%$ in the astrophysically realistic model A with $M_{a}/M_{c} \approx 1$. We also find smaller but still significant increases in $|\epsilon|$ in models B $(\approx 30 \%$ with $M_{a}/M_{c} \approx 1)$, and C $(\approx 27\%$ with $M_{a}/M_{c} \approx 1)$. {This result is consistent with the leading-order behaviour of $\epsilon$ given by \eqref{eq:empepsilon}, which implies that $|\epsilon|$ increases with $M_{c}$ \citep{mp05}. As noted in Sec. 4.1, all runs display a $\gtrsim$ two-fold increase in $M_{c}$ at $t = 2 \tC$. Thermal conduction tends to facilitate the poleward drift of matter (see Sec. 3.2), thereby making the star more prolate. }

%Mountain matter is easier to compress, when a thermal flux \eqref{eq:heatflux} is present in the simulation, supporting the conclusion in Sec. 4.1, that thermal conduction `softens' the effective EOS.

%The strong deviation in $\epsilon$ for model C with $M_{a} = 2 \times 10^{-6} M_{\odot}$ is to be expected from Fig. 7, where a region of strong density forms near the magnetic pole, which then contributes to the ellipticity.  Switching on thermal conduction appears to be most significant in the case of model C, where $\epsilon$ differs by $\sim 20 \%$ between the final states $(t = 25 \tA)$ run with and without thermal conduction for all $M_{a}$. Though not shown, the reason for this is likely because a dense region forms near the magnetic pole in all cases when thermal conduction is switched on (as in Fig. 7). This occurs due to the strong thermal flux $|\boldsymbol{F}|$ in this region (Fig. 13), which draws matter toward the magnetic pole.

The wobble angle of a precessing prolate star tends to grow, until the rotation and principal axes are orthogonal \citep{cutler02}, which is the optimal state for gravitational wave emission. Hence, an accreting neutron star with a prolate magnetic mountain may be harder to detect than an isolated magnetar with the same $|\epsilon|$, which is oblate \citep{mmra11,SMM16}. {Note that, as discussed in Sec. 2.3, the values of $|\epsilon|$ presented in this section should be treated as upper limits since we do not model sinking \citep{wette10}.}

A summary of simulation parameters and results is given in Table \ref{tab:tableparams}.

\begin{table*}
%\begin{minipage}{\textwidth}
\centering
% \begin{minipage}{160mm}

  \caption{Simulation parameters {for $\mu_{i} = B_{\star} R_{\star}^{3} \approx 3.2 \times 10^{30}$ $\text{G}$ $\text{cm}^3$.}}
  \begin{tabular}{llcccccc}
  \hline
Time & $\boldsymbol{F} \neq 0$ & EOS & $M_{a}$  & $\mu / \mu_{i}$ & $|\epsilon|$ & $\rho_{\textrm{max}}$  & $|\boldsymbol{B}_{\textrm{max}}|$ \\
 & (yes/no)  &  & $(M_{\odot})$ &  & $(10^{-8})$ & $(10^{9} \textrm{ g} \text{ cm}^{-3})$ & $(10^{12} \textrm{G})$\\
\hline
$t=0$ & --- & A & $1.2 \times 10^{-5}$ & $  0.81$ & $9.4 \times 10^{2}$ & $1.8 \times 10^{4}$ & $2.1 \times 10^{3} $  \\
   & --- & A & $1.8 \times 10^{-5}$ & $ 0.63 $ & $1.2 \times 10^{3}$ & $2.2 \times 10^{4}$ & $2.9 \times 10^{3}$  \\
   & --- & A & $2.4 \times 10^{-5}$ & $  0.49$ & $1.2 \times 10^{3}$ & $2.4 \times 10^{4}$ & $3.3 \times 10^{3}$  \\
   & --- & B & $1.0 \times 10^{-8}$ & $  0.92 $ & $0.91$ & $0.50$ & $60$ \\
   & --- & B & $2.0 \times 10^{-8}$ & $  0.73 $ & $1.3$ & $0.59$ & $98$  \\
   & --- & B & $3.0 \times 10^{-8}$ & $  0.46$ & $1.2$ & $0.64$ & $1.2 \times 10^{2}$  \\
   & --- & C & $1.0 \times 10^{-6}$ & $  0.84$ & $78$ & $3.2 \times 10^{2}$ & $6.5 \times 10^{2}$ \\
   & --- & C & $1.5 \times 10^{-6}$ & $  0.68$ & $90$ & $3.5 \times 10^{2}$ & $8.3 \times 10^{2}$  \\
   & --- & C & $2.0 \times 10^{-6}$ & $  0.50$ & $85$ & $3.7 \times 10^{2}$ & $9.7 \times 10^{2}$  \\
\hline
$t=2 \tC \gg \tA$ & N &  A & $1.2 \times 10^{-5}$ & $ 0.75 $ & $1.0 \times 10^{3}$ & $1.1 \times 10^{4}$ & $1.3 \times 10^{3} $  \\
   & N & A & $1.8 \times 10^{-5}$ & $ 0.60 $ & $1.4 \times 10^{3}$ & $1.2 \times 10^{4}$ & $1.9 \times 10^{3}$  \\
   & N & A & $2.4 \times 10^{-5}$ & $ 0.44 $ & $1.5 \times 10^{3}$ & $1.4 \times 10^{4}$ & $2.0 \times 10^{3}$  \\
   & N & B & $1.0 \times 10^{-8}$ & $ 0.86 $ & $ 0.83 $ & $0.53 $ & $50$ \\
   & N & B & $2.0 \times 10^{-8}$ & $  0.77 $ & $1.3$ & $0.61$ & $68$  \\
   & N & B & $3.0 \times 10^{-8}$ & $  0.55 $ & $1.5$ & $0.68$ & $78$  \\
   & N & C & $1.0 \times 10^{-6}$ & $ 0.91 $ & $76$ & $3.1 \times 10^{2}$ & $5.0 \times 10^{2}$ \\
   & N & C & $1.5 \times 10^{-6}$ & $  0.81 $ & $97$ & $3.4 \times 10^{2}$ & $5.9  \times 10^{2}$  \\
   & N & C & $2.0 \times 10^{-6}$ & $ 0.65 $ & $99$ & $3.5 \times 10^{2}$ & $6.5 \times 10^{2}$  \\
\hline
$t= 2 \tC \gg \tA$ & Y &  A & $1.2 \times 10^{-5}$ & $ 0.83 $ & $1.0 \times 10^{3}$ & $1.1 \times 10^{4}$ & $1.3 \times 10^{3} $  \\
   & Y & A & $1.8 \times 10^{-5}$ & $  0.73 $ & $1.5 \times 10^{3}$ & $1.3 \times 10^{4}$ & $2.0 \times 10^{3}$  \\
   & Y & A & $2.4 \times 10^{-5}$ & $ 0.65 $ & $1.7 \times 10^{3}$ & $1.5 \times 10^{4}$ & $2.1 \times 10^{3}$  \\
   & Y & B & $1.0 \times 10^{-8}$ & $  0.89 $ & $0.85$ & $0.89$ & $51$ \\
   & Y & B & $2.0 \times 10^{-8}$ & $  0.87 $ & $1.4$ & $0.82$ & $69$  \\
   & Y & B & $3.0 \times 10^{-8}$ & $  0.81 $ & $1.7$ & $0.90$ & $79$  \\
   & Y & C & $1.0 \times 10^{-6}$ & $ 0.97 $ & $78$ & $3.5 \times 10^{2}$ & $5.1 \times 10^{2}$ \\
   & Y & C & $1.5 \times 10^{-6}$ & $  0.90 $ & $1.0 \times 10^{2}$ & $3.7 \times 10^{2}$ & $6.0 \times 10^{2}$  \\
   & Y & C & $2.0 \times 10^{-6}$ & $ 0.77 $ & $1.1 \times 10^{2}$ & $4.0 \times 10^{2}$ & $6.6 \times 10^{2}$  \\

\hline
\end{tabular}
%\end{minipage}
\label{tab:tableparams}
%\end{minipage}
\end{table*}

\section{Long-Term Thermal Relaxation}

In this section we explore the long-term thermal relaxation of a representative example of an {astrophysically realistic mountain, namely model A with $M_{a} = 1.8 \times 10^{-5} M_{\odot} \approx 0.58 M_{c}$}. Long-term $(t \gg \tC)$ simulations face numerical difficulties because of the wide range of time-scales in the problem. For a typical mountain, maintaining a resolution of $128 \times 128$ grid points (see Appendix A) requires a time-step satisfying $\Delta t \lesssim 10^{-7} \tC$ to avoid numerical instabilities. It is impractical to evolve the simulation for long times $( t \gg \tC)$. Lower-resolution runs (e.g. $64 \times 64$) fail catastrophically at $t \gtrsim \tC$, because steep gradients are handled poorly at the now `blurry' mountain-atmosphere interface; one ends up with $\rho < 0$ in places, for example. To circumvent these difficulties, we artificially increase the conduction coefficients $\kappa_{\perp}$ and $\kappa_{||}$ to accelerate thermal relaxation; \cite{vigohm09} took a similar approach to accelerate Ohmic decay. Increasing $\kappa$ by a factor $\gtrsim 50$ causes the super-time-stepping algorithm to fail, when the parabolic Courant condition is eventually violated [see Appendix A and \cite{aag96}]. However, for an acceleration factor of $\lesssim 50$, the simulation is stable. %which are within two orders of magnitude of each other [FIX]. allowing for the problem to be tractable .

\begin{figure*}
\includegraphics[width=\textwidth]{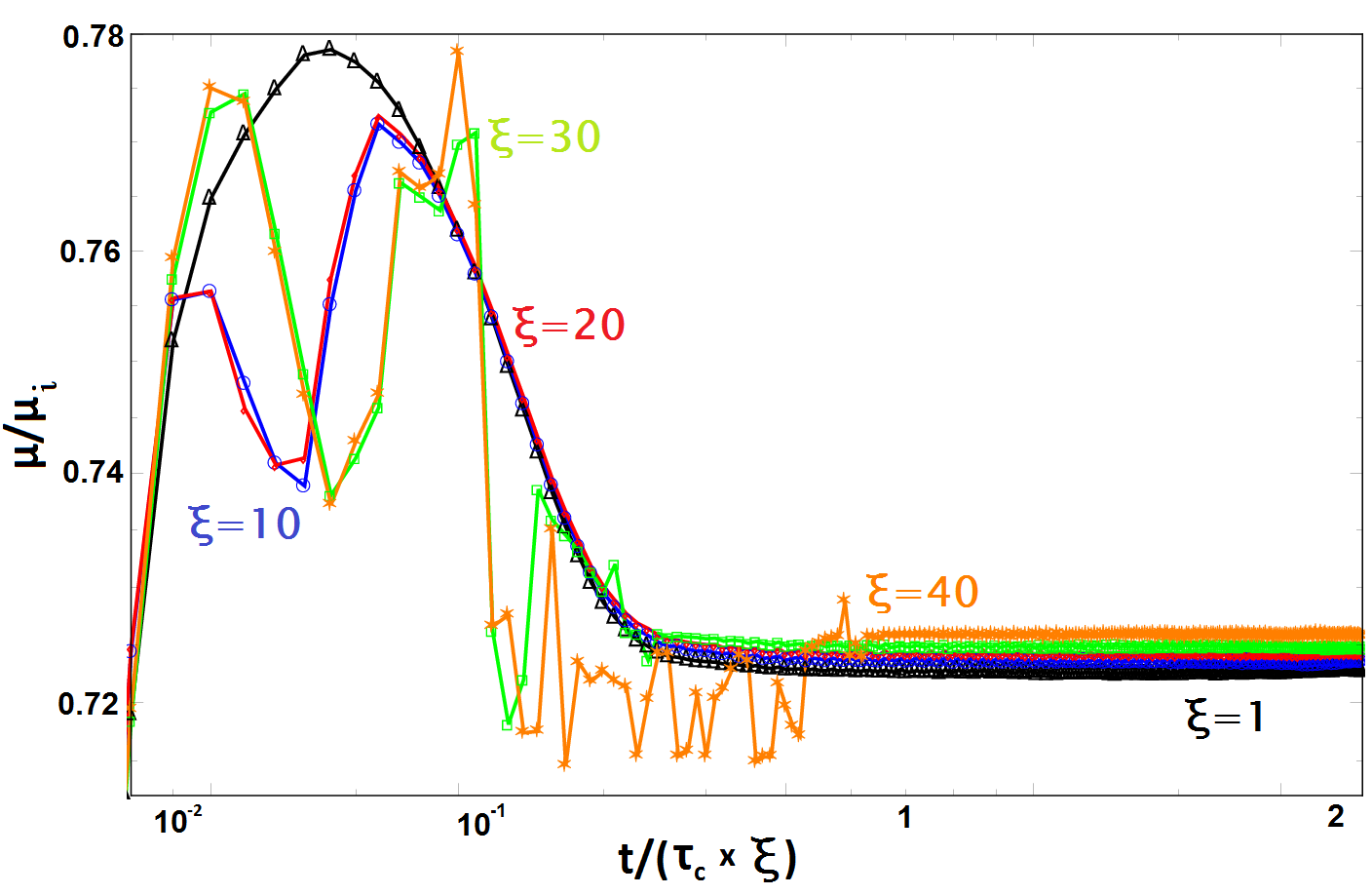}
\caption{Magnetic dipole moment $\mu$ (normalised to the pre-accretion value $\mu_{i}$) as a function of time for model A with $M_{a} = 1.8 \times 10^{-5} M_{\odot} \approx 0.58 M_{c}$ and thermal conduction switched on. The conductivites $\kappa_{\perp}$ and $\kappa_{||}$ are artificially rescaled according to $\xi \kappa_{\perp}$ and $\xi \kappa_{||}$ for $\xi = 1$ (black, triangles), $\xi = 10$ (blue, circles), $\xi = 20$ (red, diamonds), $\xi = 30$ (green, squares),  and $\xi = 40$ (orange, stars). The horizontal axis is also rescaled by expressing time in units of $\xi \tC$. \label{kappamu}
}
\end{figure*}

\begin{figure*}
\includegraphics[width=\textwidth]{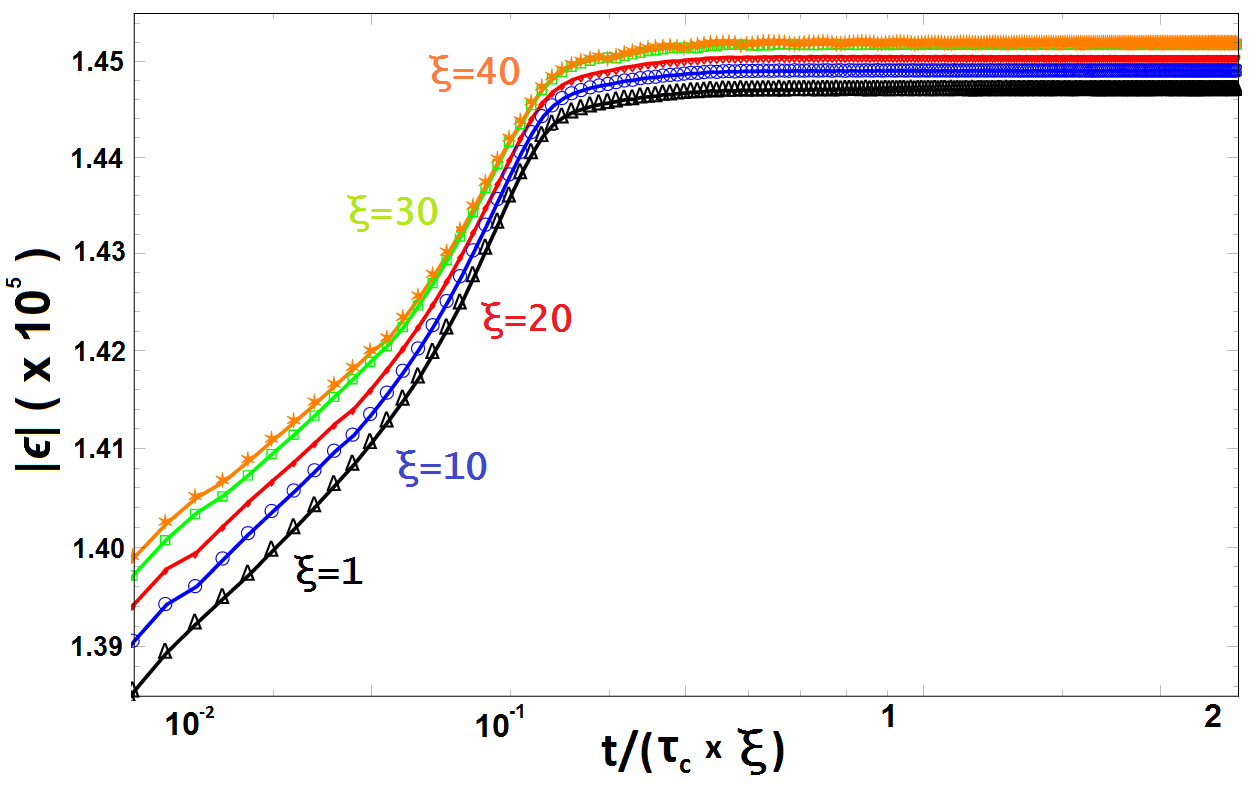}
\caption{Ellipticity $\epsilon$ [equation \eqref{eq:epsilon}] for the simulation parameters in Fig. \ref{kappamu}. Note that $\epsilon$ is negative. \label{kappaeps}
}
\end{figure*}

In order to increase the conductivities artificially, we set $\kappa_{\perp} \mapsto \xi \kappa_{\perp}$ and $\kappa_{||} \mapsto \xi \kappa_{||}$, where $1 \leq \xi \leq 40$ is a dimensionless constant. Equations \eqref{eq:alfventime} and \eqref{eq:thermaltime} imply $\tC \approx 112/\xi \textrm{ s}  \approx 5.6 \times 10^{6} \tA/\xi$. Figures \ref{kappamu} and \ref{kappaeps} display $\mu$ and $\epsilon$ respectively as functions of time for $\xi = 1,10,20,30,40$. To read the horizontal-axis for the $\xi = 30$ case, for example, a value on the axis of $10^{-1}$ implies that an interval lasting $t=  3\tC$ has effectively elapsed. The longest run effectively extends over the interval $0 \leq t \leq 80 \tC$. %e.g. $3 \tC$ for $\xi = 30$. %run the time-scale is to 

%[Sentence on how to read the $t$-scale.]

Both $\mu$ and $|\epsilon|$ increase with $\xi$. In other words, as $\kappa_{\perp}$ and $\kappa_{||}$ increase, magnetic burial is mitigated, while the gravitational wave strain increases. Thermal conduction pushes matter towards the pole (as in Sec. 3.2 and Fig. \ref{eosdenmag}). Increasing $\boldsymbol{F}$ by a factor $\xi$ amplifies polarward transport, i.e. $\rho$ increases at the pole, as $\xi$ increases, which is why $|\epsilon|$ increases with $\xi$ and the star becomes more prolate. Increasing $\rho$ near the pole effectively reduces the fraction $p / \rho^{\Gamma} = k_{\Gamma}$ there; i.e. increasing $\xi$ can be thought of as reducing the effective polytropic constant and `softening' the EOS by a factor related to $\xi$; cf. Table \ref{tab:table1eos}. Hence, initially adiabatic mountains evolved with high $\boldsymbol{F}$ come to resemble isothermal mountains at $t \gg \tC$ [compare Fig. \ref{kappamu} with Figure 8 of \cite{pm04}]. For example, for $\xi = 40$, we have $\mu(t \gg \tC)/\mu_{i} = 0.73$, cf. $\mu(t=0)/\mu_{i} = 0.63$. By comparison, we find $\mu/\mu_{i}= 0.73$ and $\mu/ \mu_{i} = 0.63$ from isothermal Grad-Shafranov simulations {(softest EOS)} for $M_{a} \approx 10^{-5} M_{\odot}$ and $M_{a} \approx 10^{-4} M_{\odot}$ respectively. Comparing the $\xi = 1$ and $\xi = 40$ final states, we find that $\mu$ and $\epsilon$ differ by $\leq 3\%$ and $\leq 1\%$ respectively. {The trends discussed above are also evident in simulations with different EOS and grid resolutions; see Appendix A.}

%This is reminiscent of what occurs for a softer equation of state, such as for an isothermal EOS (model A), where $M_{c}$ is large and a greater accreted mass is required to bury the dipole moment; cf. equation \eqref{eq:empirical}.
%This behaviour may be explainable, through relation \eqref{eq:estimate2}, increasing $\kappa_{||}$ by a factor $\xi$ is equivalent to increasing $T_{e}$ by a factor $\xi^{2/5}$ when $\kappa_{\perp} \ll \kappa_{||}$. Since $T_{e}$ is proportional to the polytropic constant $K_{\Gamma}$ through \eqref{eq:temp}, we see that increasing $\kappa$ has the same effect as increasing $K_{\Gamma}$ and hence `softens' the effective EOS by a factor $\xi^{2/5}$ (cf. Table \ref{tab:table1eos}). 
 %Overall, we conclude that the long-term thermal relaxation of the mountain does not significantly alter the global observables.

%[Re this and explain oscillations.]
We see that $\mu$ and $\epsilon$ continue to oscillate after $t \gtrsim 2 \tC$ albeit with small amplitude $(\lesssim 1\%$ peak to peak$)$. The fluctuations at $t \gg \tC$ persist, because some parts of the mountain take longer to settle down than others. In reality, heat transport occurs more slowly than average in cold regions ($T_{e} \ll 10^{8} \text{ K}$), meaning that conduction continues to affect cold parts of the mountain (whose effective conduction time-scales are longer than the volume-averaged value $\tC$), even after the rest of the mountain relaxes thermally. These cold regions, however, do not play a dominant role in determining $\mu$ or $\epsilon$, as the density is low there. %but nevertheless can be seen meaning the mountain as a whole continues to oscillate slightly beyond $t \gg \tC$ using definition \eqref{eq:thermaltime}. 
%In any case, however, minimal evolution is seen after $t \gtrsim \tC$ s for any $\xi$. %indicating perhaps that the estimate \eqref{eq:thermaltime} is, in general, an over-estimate.

%The time-scales observed in the run are somewhat conflicting. From Fig. 21 it is clear that the $\xi=1$ run takes longer to settle to its equilibrium value of $\epsilon$, as can seen from the oscillations in the black curve. The runs stabilise earlier with increasing $\xi$, indicating that the thermal time-scale is shortening with increasing $\xi$ as is expected from \eqref{eq:timeest}. The dipole moment in Fig. 20, however, appears to stabilise at the same value of $t$ independently of the value of $\xi$. This is suggesting that the magnetic field is unperturbed beyond the dynamical time-scale regardless of the numerical value of the conduction coefficient, while the density profile is sensitive to the time-scale \eqref{eq:taucestimate}. It appears that the energy equation \eqref{eq:energy} saturates in $\bb$ but not in $\rho$ therefore; though further investigation into the energy equation \eqref{eq:energy} for large $\kappa$ would be required to conclude this in general.

\section{Conclusions}

In this paper we explore the effects of thermal conduction on the evolution of accretion-built mountains on neutron stars for time-scales $t \gtrsim \tC$ (Secs. 3 and 4) and $t \gg \tC$ (Sec. 5) using the MHD code \PLUTO \,\citep{pluto}. The initial states are generated from the Grad-Shafranov equation for a range of initially polytropic EOS documented in Table \ref{tab:table1eos}  \citep{pm04,pri11}. Model A approximates a realistic, non-catalysed, accreted crust for densities in the range $10^{8} \leq \rho / \text{ g cm}^{-3} \leq 10^{14}$ \citep{haen90}. Models B and C approach the realistic EOS in the low-$\rho$ and neutron drip density regimes respectively, and are included for completeness to illustrate EOS-related trends and make contact with previous work. %We explpre the conduction-altered relationship between the accretion-induced magnetic burial and the gravitational radiation associated with the mountain quadrupole moment. 
The theory of magnetic burial predicts that, as matter piles up on the stellar surface, the dipole moment is reduced in accreting neutron stars in accord with the observed $\mu$ versus $M_{a}$ relations, e.g. \cite{taam86,hue95}. We find that thermal conduction has the effect of pushing accreted matter back towards the magnetic pole, {where $\bsn T$ is greatest}, thereby partially resurrecting the buried field and increasing $\mu$ while making the star more prolate. On the conduction time-scale, we find a quasi-static increase in the mountain's characteristic mass $M_{c}$ [defined above equation \eqref{eq:empirical}] starting from an adiabatic initial state. %, and a decrease in $M_{c}$ starting from an isothermal initial state. %These results suggest that thermal conduction `softens' the effective EOS for initially adiabatic EOS, while the reverse occurs for isothermal mountains.
Hence achieving a given $\mu/\mu_{i}$ value requires higher $M_{a}$, in general, than estimated by \cite{pri11}. The main trends are summarised in Table \ref{tab:tableparams}.

Gravitational radiation back-reaction can stall the spinup of the neutron star in a low-mass X-ray binary (LMXB) at hectohertz frequencies \citep{bildstenpaper}, explaining the observation that LMXBs spin slower $(\nu_{s} \lesssim 650 \text{ Hz})$ than otherwise expected \citep{chaknat03}. The results in this paper suggest that the effective EOS of mountain matter may be softer than previously estimated, when thermal conductivity is included, leading to a proportionally higher gravitational wave strain \eqref{eq:wavestrain}. This strengthens the argument for targeting LMXBs such as Sco X-1 for searches with facilities like LIGO \citep{scox1,riles13,hask15}. The increase in $|\epsilon|$ combined with the decrease in $\mu/\mu_{i}$ at $t \gtrsim \tC$ for all runs with $M_{a}/M_{c} \gtrsim 0.1$ performed in this paper suggests that stars with significantly buried $(\mu \ll \mu_{i})$ magnetic fields may prove better gravitational wave candidates than previous estimates indicated \citep{mp05,pri11}. {However, we stress that the systematic and numerical (see Appendix A) uncertainties present within our models suggest that the effects of thermal conduction are likely to be small compared to other physical effects not implemented here, such as sinking \citep{wette10}.}

In addition to searching for gravitational waves and measuring the global dipole moment, one can test the magnetic burial scenario by studying type I X-ray bursts \citep{stroh03,cumm04,pmtherm,gal08}. For adiabatic initial states, we find that mountains develop hot $(T \gtrsim 10^{9} \text{ K})$ spots with large thermal fluxes near both the pole $(\theta = 0.1)$ and the equator $(\theta = 1.3)$ for a wide range of accreted masses (see Table \ref{tab:tableparams} and Fig. \ref{eostempflux}). Dense filamentary regions also develop, especially for $M_{a}/M_{c} \gtrsim 1$. These effects cooperate to produce localized hot patches ``fenced off" by intense magnetic fields, whose number increases with $M_{a}$ [cf. \cite{heyl03}]. The hot spots may individually provide fuel for type I X-ray bursts which do not spread across the entire stellar surface, if the magnetic fences are intense enough to inhibit cross-field thermal transport \citep{gal10,misanovic10}. X-ray observations of significant heat fluxes near the magnetic pole of a neutron star in an LMXB, as broadly predicted by our simulations, may be related to the magnetic mountain physics \citep{heyl03,bhat06,watts17}. Thermal fluxes out of the hot spots may also amplify shear stresses felt by the neutron star crust \citep{chug10,belo14}. A detailed analysis of hot-spot phenomena and their observational consequences will be conducted in future work. {Another avenue to probe accretion mound physics comes from cyclotron features \citep{dip12}. \cite{pri14} showed that one can discriminate, in principle, between magnetic mountain properties (e.g. EOS) by studying the line energy, width, and depth of theoretical cyclotron resonant scattering features from accreting neutron stars. These cyclotron features are, however, unlikely to be detected in the near future as it requires further development of sensitive X-ray polarimeters \citep{hask15}.} % . [Sentence on thermo-plastic waves and extensions.] %Cooling and sunspot activity along the lines of [Refs] will be investigated in future works.

%. We find that the abundance of these hot patches is proportional to $M_{a}/M_{c}$, and thus we predict there should be a correlation between accreting systems with
%Hot spot discussion. We see a big heat flux at the pole in general adiabatic runs. This means that ...

%Reiterate cyclotron line and Arons problems?


%\acknowledgments

\section*{acknowledgments}
We thank Maxim Priymak and Donald Payne for permission to modify and use the Grad-Shafranov solver. We thank Dipanjan Mukherjee for expert instruction on the use of \PLUTO. We thank HoChan Cheon for designing an early version of the script that converts Grad-Shafranov output into a format suitable for \textsc{\footnotesize{PLUTO}} input. We thank Patrick Clearwater, Brynmor Haskell, and Alpha Mastrano for discussions. {We thank the anonymous referee for their carefully considered comments, and for providing the Skyrme EOS data used in Sec. 2.3 to estimate sinking depths.} This work was supported in part by an Australian Postgraduate Award, the Albert Shimmins fund, and an Australian Research Council Discovery Project grant. 

%[FIX: figure labels in appendix].

%%%%%%%%%%%%%%%%%%%%%%%%%%%%%%%%%%%%%%%%%%%%%%%%%%%%%%%%%%%%%%%%%%%%%%%%%%%%%%%%%%%%%%%

%%%%%%%%%%%%%%%%%%%%%%%%%%%%%%%%%%%%%%%%%%%%%%%%%%%%%%%%%%%%%%%%%%%%%%%%%%%%%%%%%%%%%%%
\appendix
%%%%%%%%%%%%%%%%%%%%%%%%%%%%%%%%%%%%%%%%%%%%%%%%%%%%%%%%%%%%%%%%%%%%

\section{PLUTO simulations}

Complete documentation for the \PLUTO \,code was published by \cite{pluto}. The specific features we rely upon and optimize are discussed below.

\subsection*{Grid and time step}

We employ a static, two-dimensional, polar grid with  $N_{r} \times N_{\theta} = 128 \times 128$ grid points. The radial grid comprises a logarithmic section with $100$ points for $R_{\textrm{in}} \leq r \leq 2 R_{\textrm{mountain}}$, and a uniform section with $28$ points for $2 R_{\textrm{mountain}} \leq r \leq R_{\textrm{m}}$, where $R_{\textrm{mountain}}$ is defined arbitrarily at $t=0$ as the innermost radial grid point with $\rho \leq 10^{-7} \rho_{\text{max}}$. A mixed grid captures features with sharply different length-scales and minimizes the interpolation errors discussed in Sec. 3. We find that including additional grid points in the atmosphere ($R_{\textrm{mountain}} \leq r \leq R_{\textrm{m}}$) increases the computational cost without modifying perceptibly the observables computed in Sec. 4. The angular grid is uniformly spaced in $\theta$ over $0 \leq \theta \leq \pi/2$.

We employ a Runge-Kutta third-order time-stepper for safety, although we find by experimentation that the results are essentially indistinguishable from the second-order variant. We employ the third-order finite-volume spatial integrator `Lim03' to interpolate between grid points \citep{lim03}. This scheme resolves local minima with high precision, e.g. strong gradients at the mountain-atmosphere interface. We use a time-step $\Delta t = 5 \times 10^{-8} \tC$ where $\tC$ is determined through equation \eqref{eq:thermaltime}. We print output files at various fractions of $\tC$ depending on the specifics of the run; cf. the horizontal-axes on Figs. \ref{eosdipmom} and \ref{eosepsilon}. This $\Delta t$ is small enough to avoid Courant-Friedrichs-Lewy (CFL) instabilities for each run.

In order to avoid numerical instabilities we simulate the mountain together with an atmosphere which has a small but non-zero density taken as $\rho_{\textrm{atm}} = 10^{-8} \rho_{\textrm{max}}$. The atmosphere alleviates numerical difficulties associated with strong gradients and prevents the density from dipping below zero due to numerical fluctuations. We find that varying $\rho_{\textrm{atm}}$ in the range $10^{-9} \leq \rho_{\textrm{atm}}/ \rho_{\textrm{max}} \leq 10^{-5}$ does not modify the observables discussed in Sec. 3 and 4 by more than $\sim 1\%$. The code crashes for $\rho_{\textrm{atm}} \lesssim 10^{-9} \rho_{\textrm{max}}$, because $\rho$ dips below zero somewhere unless we set $\Delta t \lesssim 10^{-11} \tC$, which is too expensive computationally. We define a flag that sets $\max(\rho,\rho_{\text{atm}}) \mapsto \rho$ at every grid point after each time step $\Delta t$ so that the atmosphere has a minimum density of $\rho_{\text{atm}}$ for all $t$.

We find that the bilinear interpolation algorithm of \PLUTO \,introduces errors of $\leq 1 \%$ in the quantities calculated in Secs. 3 and 4 between the $t=0$ (grid-realigned) \PLUTO \,output and the raw Grad-Shafranov data; cf. Figures 2 and 4 of \cite{pri11} with Figs. \ref{eosdipmom} and \ref{eosepsilon}.

\subsection*{Divergence cleaning and thermal conduction}

Maxwell's equations require $\bsn \cdot \bb = 0$ at all times. Various strategies can be employed to minimise  numerical deviations from $ \bsn \cdot \bb = 0$. For example, the extended hyperbolic divergence cleaning algorithm \citep{hdivcl} introduces Lagrange multipliers into Faraday's law \eqref{eq:faraday}. Inspection of \PLUTO \,output files confirms that $\bsn \cdot \bb$ vanishes to floating-point precision as a consequence of using this algorithm. The divergence cleaning algorithm is coupled with the approximate Riemann solver `hlld' designed to resolve shocks and strong gradient phenomena \citep{MK05}.

Thermal conduction (see Sec. 3) is implemented via the super-time-stepping algorithm available in \PLUTO \,and described in \cite{aag96}. The energy equation \eqref{eq:energy} has a parabolic Courant number $C_{p}$ associated with it, which depends on the value of the conduction coefficients $\kappa_{||}$ and $\kappa_{\perp}$. Together with the usual Courant number condition \citep{LL59}, we require $C_{p} \leq 1/N_{\text{dim}} = 1/2$ to avoid instabilities \citep{bec92}. Super-time-stepping allows for flux terms to be treated in a separate `super-step' using operator splitting methods, so that $\Delta t$ need not be reduced to avoid parabolic CFL instabilities.

\subsection*{Implementation, stability, and convergence tests}

\begin{figure}
%\centering
\includegraphics[width=0.473\textwidth]{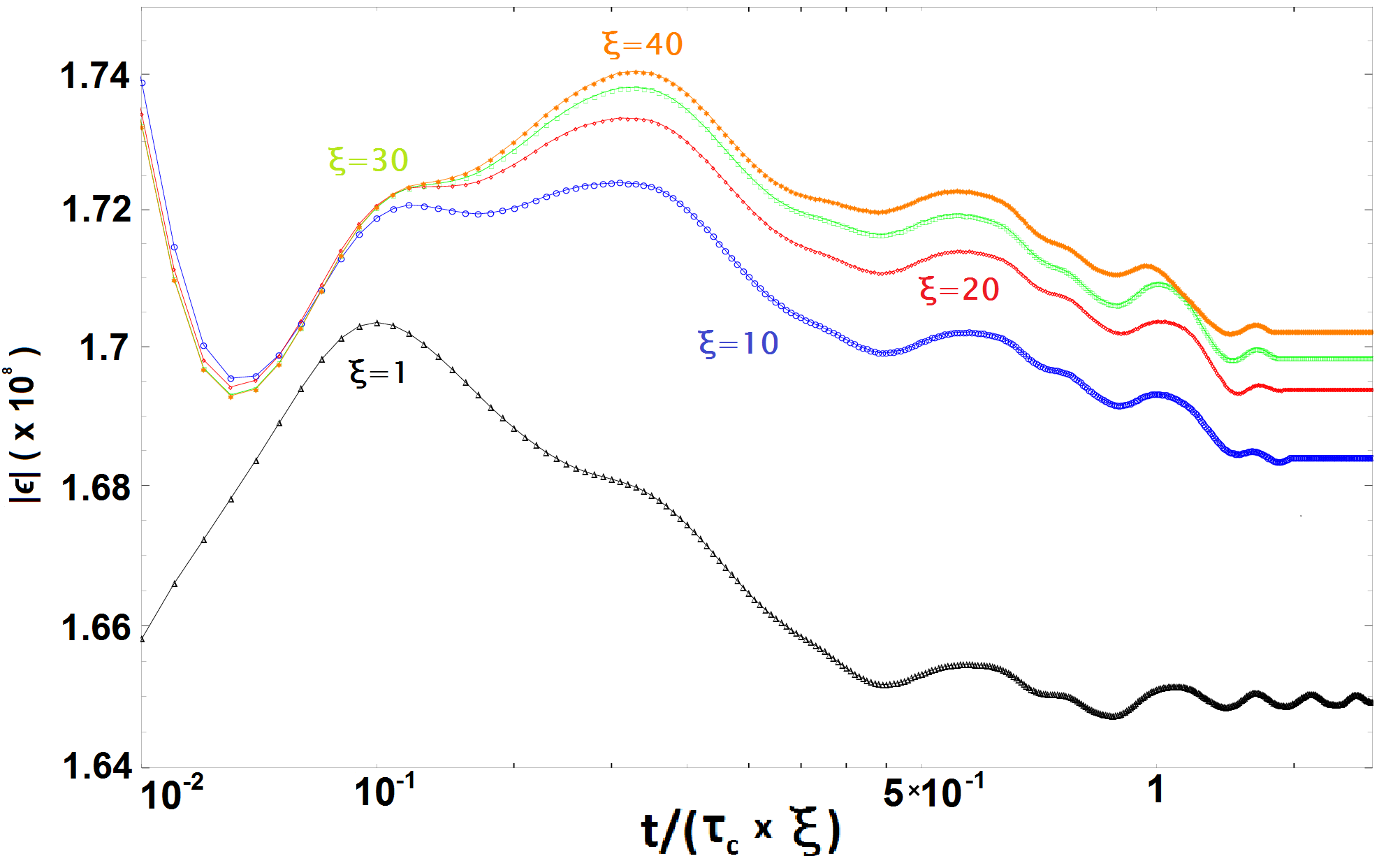}
\caption{Evolution of $\epsilon$ for model B with $M_{a} = 2 \times 10^{-8} M_{\odot}$ with $96 \times 96$ grid points for $\xi = 1$ (black, triangles), $\xi = 10$ (blue, circles), $\xi = 20$ (red, diamonds), $\xi = 30$ (green, squares), and $\xi = 40$ (orange, stars). \label{modbeps}
}
\end{figure}

\begin{figure}
%\centering
\includegraphics[width=0.473\textwidth]{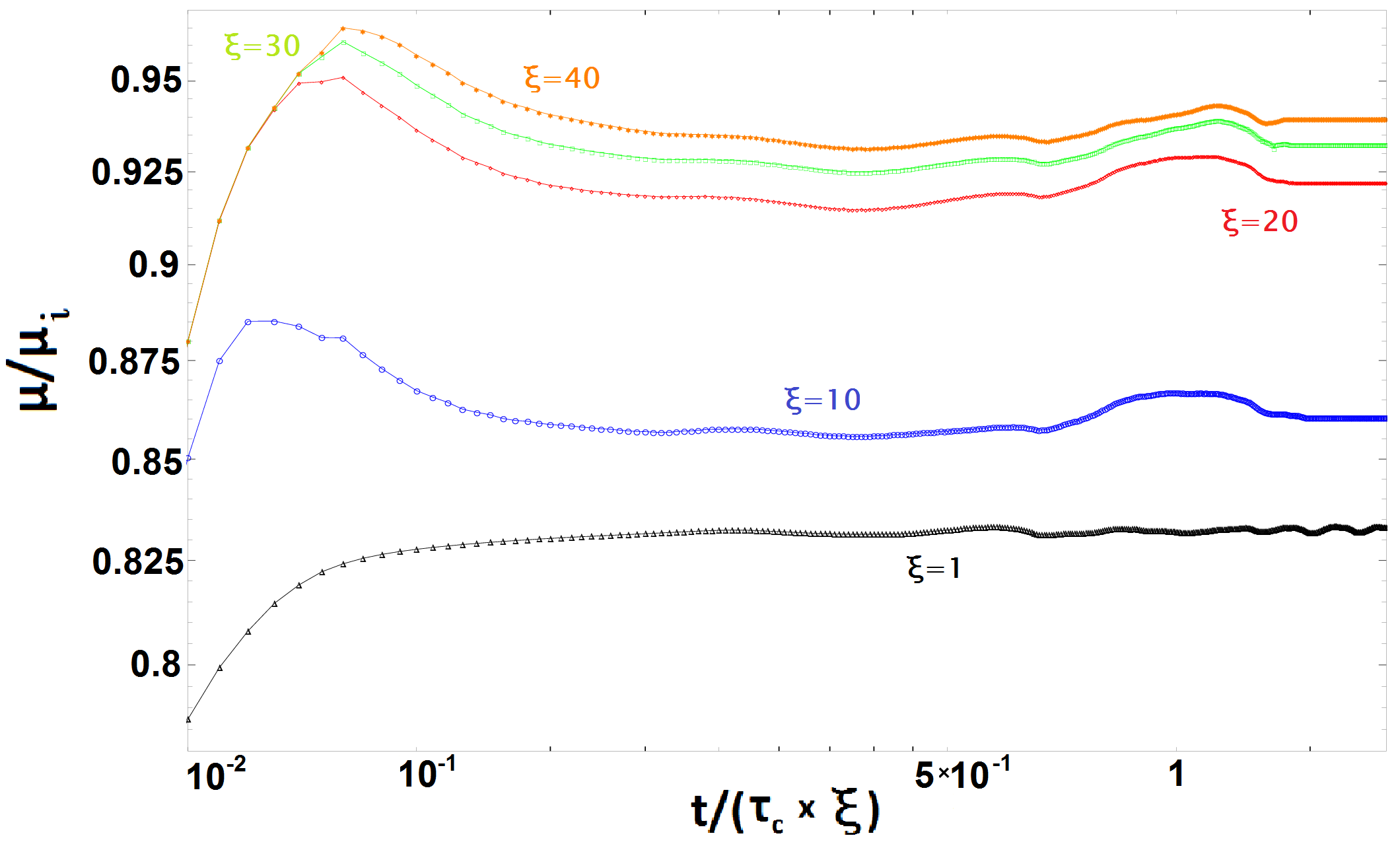}
\caption{Evolution of $\mu$ for model B with $M_{a} = 2 \times 10^{-8} M_{\odot}$ with $96 \times 96$ grid points for $\xi = 1$ (black, triangles), $\xi = 10$ (blue, circles), $\xi = 20$ (red, diamonds), $\xi = 30$ (green, squares), and $\xi = 40$ (orange, stars).  \label{modbmu}
}
\end{figure}

We test our \PLUTO\, simulations in four ways. {For implementation:} (i) We compute the total mass of the simulation at each time-step to check for mass leakage. (ii) We check the surface dipole moment and the velocity field to ensure that the boundary conditions described in Sec. 2.2 are implemented faithfully. {For convergence:} (iii) We vary the grid parameters $N_{r}$ and $N_{\theta}$ to check if the results depend on the spatial resolution {(see Figs. \ref{modbeps} and \ref{modbmu})}. {For stability:} (iv) We vary the super-time-stepping parameters, the CFL parameters, and the thermal conduction coefficients (i.e. checking if the conduction and no-conduction runs match smoothly in the limit $\kappa \rightarrow 0$).  The convergence of the Grad-Shafranov code described in Sec. 2.2 is studied fully by \cite{pm04} and \cite{pri11}. %Overall, we find that runs performed with any small modifications to the parameters presented here do not differ fundamentally from the presented results.

{In Figures \ref{modbeps} and \ref{modbmu} we show the evolution of the ellipticity and dipole moment, respectively, for model B with $M_{a} = 2 \times 10^{-8} M_{\odot}$ and thermal conduction switched on, with $N_{r} \times N_{\theta} = 96 \times 96$ grid points and varying values of $\xi$ (this parameter is introduced to artificially scale the conduction coefficients, see Sec. 5). Two major points are evident from these plots. First, the trends associated with increasing $\xi$ for model B are the same as was observed for model A in Sec. 5; $|\epsilon|$ and $\mu/\mu_{i}$ are monotonically increasing with increasing values of $\xi$ at $t \gtrsim \tau_{c}$, independent of the EOS and grid resolution. The second point concerns the convergence test (iii) detailed above: for the $\xi = 1$ run (black, diamonds), all simulation parameters are identical to those for the simulations performed in Sec. 4 for the same accreted mass and EOS (middle figure of the middle panel), except that the resolution is lower for the runs presented here. Comparing the final ellipticity and $\mu$ values from Figs. \ref{modbeps} and \ref{modbmu} with those presented in Table 3 for the higher resolution run, we see only a small ($\lesssim 10 \%$) disparity at late times, with $\epsilon(96 \times 96) = 1.65 \times 10^{-8}$, $\epsilon(128 \times 128) = 1.44 \times 10^{-8}$, $\mu(96 \times 96)/\mu_{i} = 0.83$, and $\mu(128 \times 128)/\mu_{i} = 0.87$.
}

\begin{figure}
\includegraphics[width=0.473\textwidth]{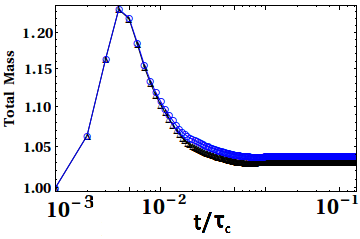}
\caption{Total mass enclosed in the simulation box (in units of $M_{a}$) as a function of time for model B with $M_{a} = 2 \times 10^{-8} M_{\odot}$, evolved with (blue, circles) and without (black, triangles) thermal conduction. \label{masstest}
}
\end{figure}

Figure \ref{masstest} plots the total mass as a function of time without (black, triangles) and with (blue, circles) thermal conduction for model B with $M_{a} = 2 \times 10^{-8} M_{\odot}$. We see that, after an initial adjustment phase, the total mass returns to $M_{a}$ within $\sim 3 \%$ ($\sim 4 \%$) without (with) thermal conduction. {This adjustment phase $(t \ll \tau_{c})$ occurs for two separate reasons. The first is due to the artificial atmospheric density $\rho_{\text{atm}} = 10^{-8} \rho_{\text{max}}$, introduced to ensure that the simulation does not produce $\rho < 0$ at any point throughout the evolution. Some of this atmospheric mass actually gets pulled down into the mountain, after which the atmosphere resets, thus increasing the overall mass of the simulation slightly. Additionally, the Grad-Shafranov equilibria are defined over grids which are slightly different to those in PLUTO. Hence, at $t=0$, the MHD equations are not exactly satisfied in PLUTO, leading to a temporary increase in the total mass. These two effects combine to increase the total mass in the initial stages of evolution. Table A2 in \cite{pstab07} reports similar total mass changes during the adjustment phase.} Figure \ref{masstest} is typical for runs performed in this paper.

\begin{figure}
\includegraphics[width=0.473\textwidth]{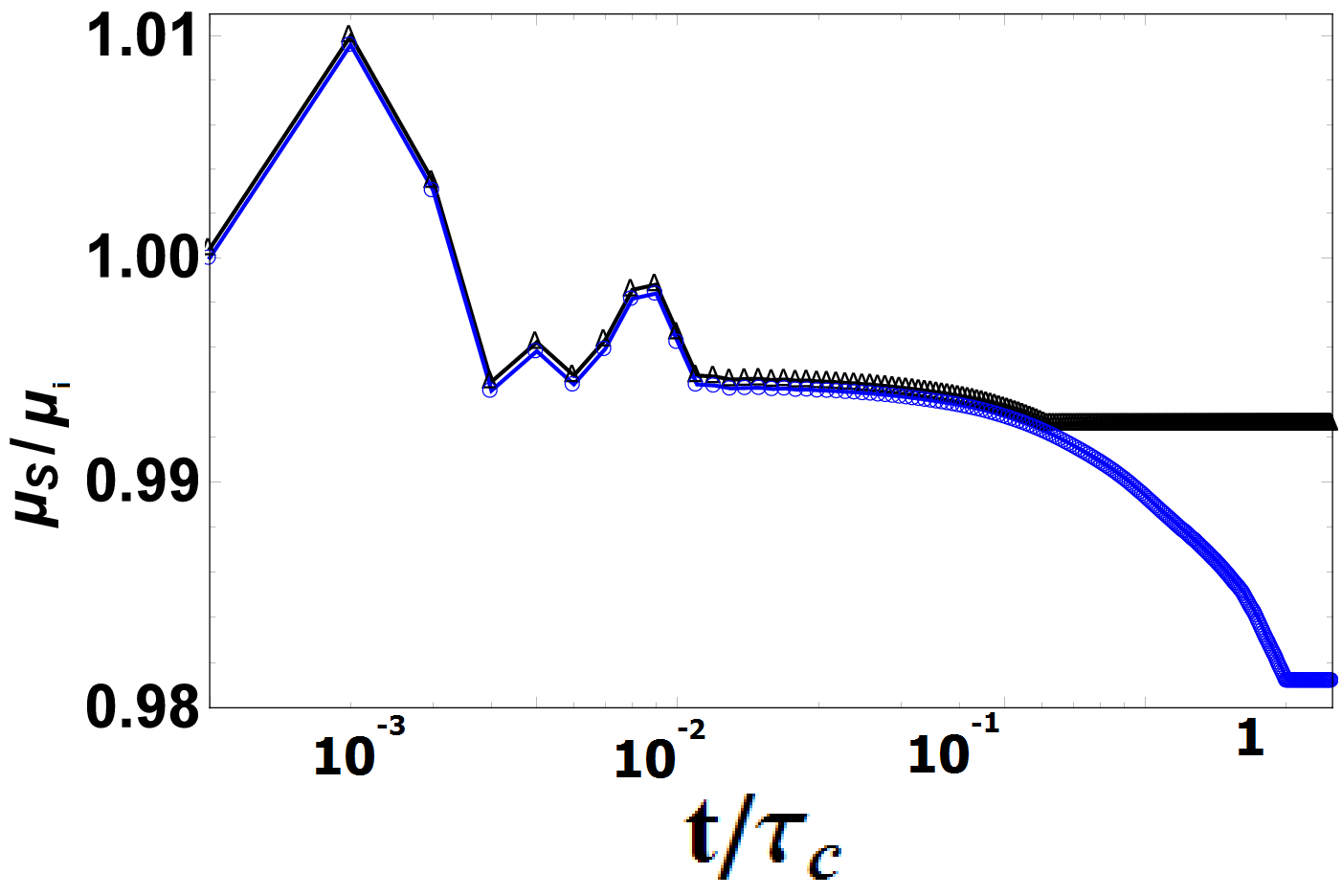}
\caption{Surface magnetic dipole moment $\mu_{\text{S}}$ [equation \eqref{eq:dipmom} evaluated at $r = R_{\textrm{in}}$] as a function of time, normalised to the pre-accretion value $\mu_{i}$ for model A with $M_{a} = 1.8 \times 10^{-5} M_{\odot}$, evolved with (blue, circles) and without (black, triangles) thermal conduction. \label{dipmomtest}
}
\end{figure}

In Figure \ref{dipmomtest} we plot the surface dipole moment $\mu_{\text{S}}$ [equation \eqref{eq:dipmom} evaluated at $r = R_{\textrm{in}}$] as a function of time without (black, triangles) and with (blue, circles) thermal conduction for model A with $M_{a} = 1.8 \times 10^{-5} M_{\odot}$. If the boundary conditions at the stellar surface [namely $\psi(R_{\text{in}},\theta) = \psi_{\star} \sin^2\theta$] are implemented without numerical error, $\mu_{\text{S}}$ should keep its initial value $\mu_{i}$. We see a slight variation (maximum of $\sim 2 \%$). Figure \ref{dipmomtest} is typical for runs performed in this paper.

\section{Ideal-gas approximation to the equation of state}

Strictly speaking, the accreted matter in the crust is partially degenerate \citep{schatz99}. In this appendix, we verify that it is reasonable to approximate the EOS by the ideal-gas formula \eqref{eq:temp}, for ease of use in \PLUTO, when calculating the perturbations to the mountain structure caused by thermal transport. The equilibrium configuration of the mountain before thermal transport is switched on is calculated for the full, degenerate, polytropic EOS (see Sec. 2).

In a Fermi-Dirac distribution, the mean occupancy $n$ for a single-particle orbital with energy $E$ is given by
\begin{equation} \label{eq:states}
n(E) = \frac {1} {\exp\left[{\frac {E - \sigma} {k_{B} T}}\right] +1 },
\end{equation}
where $\sigma$ is the chemical potential, which is a function of $\rho$ and $T$ (in general). In the limit $T \rightarrow 0$, $n(E)$ tends to either $1$ or $0$ for $E < \sigma$ or $E > \sigma$, respectively. The Fermi temperature $T_{F}$ is defined through the chemical potential via
\begin{equation}
\sigma = k_{B} T_{F}.
\end{equation}
The dependence of $\sigma$ on $\rho$ and $T$ is determined by integrating the mean occupancy to obtain the total particle number,
\begin{equation} \label{eq:denint}
N = \int^{\infty}_{0} n(E) D(E) d E,
\end{equation}
where $D(E)$ is the density of states.

\begin{figure}
\includegraphics[width=0.473\textwidth]{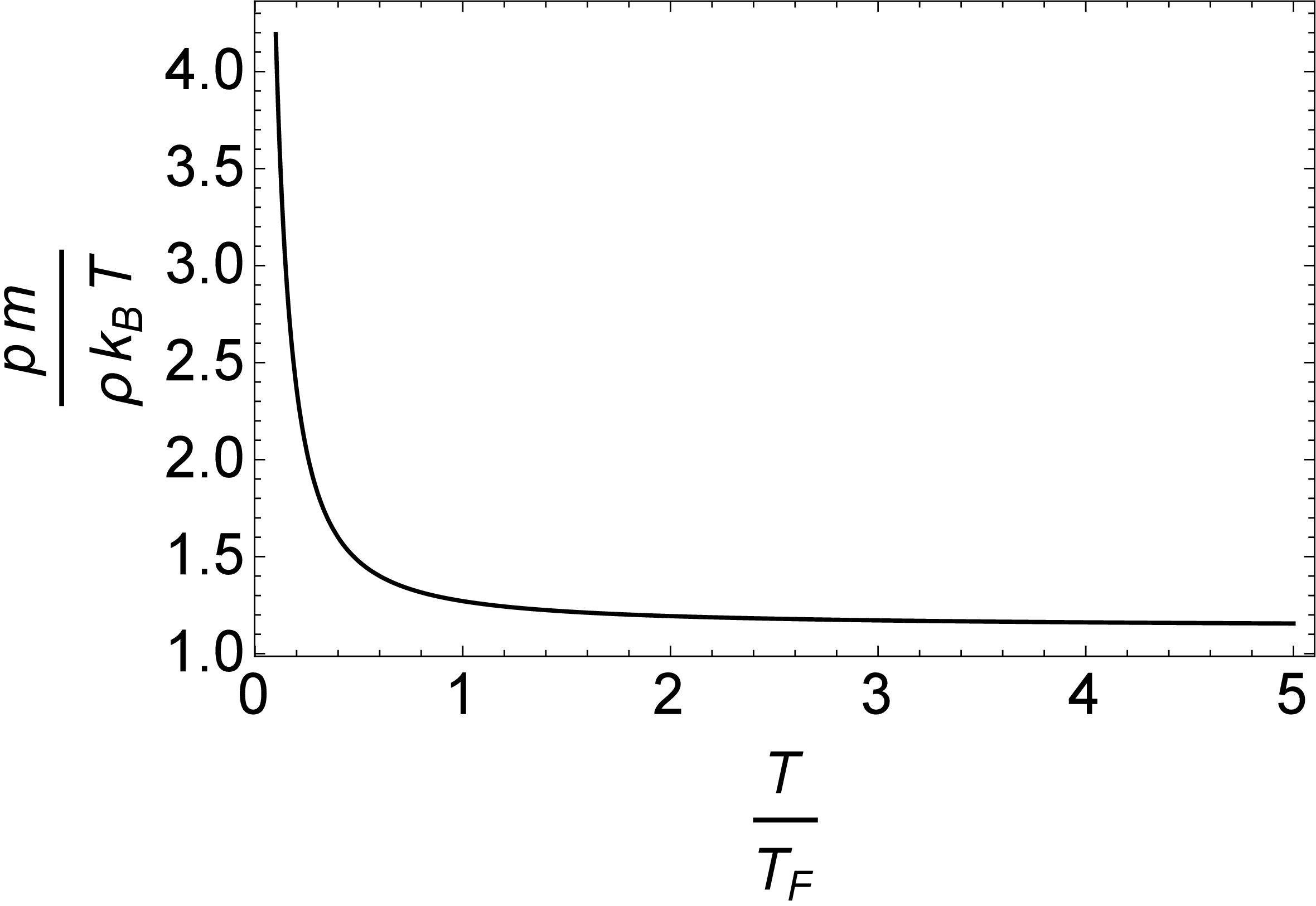}
\caption{Thermal EOS for a nonrelativistic Fermi gas. \label{figXstar}}
\end{figure}

The pressure $p$ is defined via the first law of thermodynamics, viz. 
\begin{equation} \label{eq:grand}
p =  -\frac {\left(T - \varepsilon S - \sigma N \right)} {V},
\end{equation}
where $\varepsilon$ and $S$ denote the internal energy and entropy, respectively, and $-pV$ is the grand canonical potential. Substituting \eqref{eq:grand} into the integral \eqref{eq:states} allows one to express $T$ in terms of $p$ and $\rho$ for $T \neq 0$, i.e. defines the EOS. One finds
\begin{equation} \label{eq:realeos}
\frac {p m} {\rho k_{B} T} = \frac {\mathcal{F}(5/2,z)} {\mathcal{F}(3/2,z)},
\end{equation}
for an arbitrary Fermi gas, with
\begin{equation}
\mathcal{F}(\nu,z) = \frac {1} {\Gamma(\nu)} \int^{\infty}_{0} \frac {x^{\nu-1}} {z^{-1} e^{x} +1 } dx,
\end{equation}
and fugacity $z = \exp{\left( {\sigma} / {k_{B} T} \right)}$ [see e.g. \cite{ST83} for details]. Expression \eqref{eq:realeos} is plotted in Fig. \ref{figXstar}.

In regions with $T \gg T_{F}$, expression \eqref{eq:realeos} approaches the ideal gas law \eqref{eq:temp},
\begin{equation} \label{eq:formr}
p \propto \rho k_{B} T.
\end{equation} 
In regions with $T \ll T_{F}$, expression \eqref{eq:realeos} approaches a polytropic EOS \citep{ch76},
\begin{equation} \label{eq:lattr}
p \propto \rho^{5/3}.
\end{equation}
An accurate description of a realistic accreted crust lies between these two extremes \citep{schatz99}. The latter limit \eqref{eq:lattr} coincides with the Grad-Shafranov $t=0$ initial condition for degenerate, single-particle fluids, e.g. models B and C in Table \ref{tab:table1eos} and Figure \ref{zdunik}. The Grad-Shafranov equilibria, calculated with \eqref{eq:lattr}, adjust modestly, when thermal transport is switched on in \PLUTO\, with \eqref{eq:formr}, suggesting that the equilibrium starting-point is broadly consistent with both \eqref{eq:formr} and \eqref{eq:lattr}, for the values of $\kappa_{||}$ and $\kappa_{\perp}$ relevant here.

As noted throughout the body of the paper, it is the hot regions of the mountain where thermal conduction modifies the hydrodynamic structure the most as time passes. This is expected because the (dominant) parallel thermal conductivity $\kappa_{||}$ scales as $\kappa_{||} \propto T^{5/2}$ through \eqref{eq:estimate1}. In hot regions, we have $T \gtrsim T_{F}$. Hence Fig. \ref{figXstar} implies a $\lesssim 20\%$ departure in $p m/ \rho k_{B} T$ from the ideal gas law.
 
\begin{figure*}
\includegraphics[width=\textwidth]{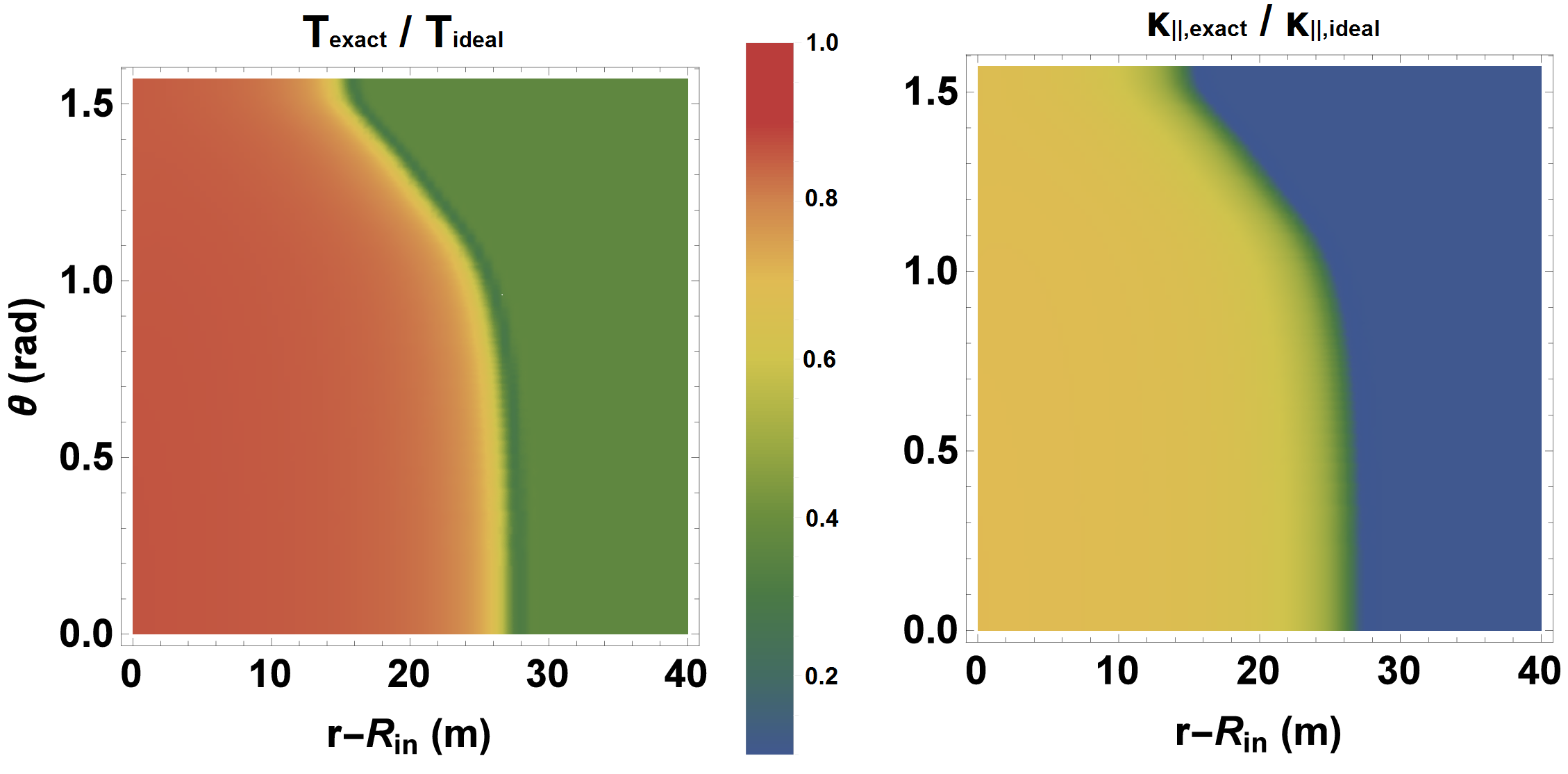}
\caption{Contours of $T_{\text{exact}}$ [computed using expression (B5)] normalised by $T_{\text{ideal}}$ [computed using expression \eqref{eq:temp}] (left panel) and the similarly defined parallel conduction coefficients ratio $\kappa_{||,\text{exact}} / \kappa_{||,\text{ideal}}$ (right panel), plotted at $t=0$ for realistic crust model A with $M_{a} = 1.8 \times 10^{-5} M_{\odot} \approx 0.58 M_{c}$. Red shades indicate values close to unity, while blue shades indicate values close to 0.1. \label{tempcomp}}
\end{figure*}

In Figure \ref{tempcomp} we plot contours of $T_{\text{exact}}/T_{\text{ideal}}$ [i.e. $T_{\text{exact}}$ from \eqref{eq:realeos} divided by $T_{\text{ideal}}$ from \eqref{eq:temp}] (left panel) and $\kappa_{||,\text{exact}} / \kappa_{||,\text{ideal}}$ (similarly defined, right panel) for the realistic accreted crust model A with $M_{a} = 1.8 \times 10^{-5} M_{\odot} \approx 0.58 M_{c}$ at $t=0$. We find $T_{\text{exact}}/T_{\text{ideal}} \gtrsim 0.85$ throughout the bulk of the mountain, i.e. the temperature is overestimated by $\lesssim 15 \%$ in the densest regions of the mountain, where most mass resides, for this representative simulation. This translates into a $\lesssim 40 \%$ overestimate in $\kappa_{||} \gg \kappa_{\perp}$ everywhere except at the mountain-atmosphere interface, where there is little mass, and the model breaks down anyway {because of the artificial $\rho_{\text{atm}}$}.

\end{document}